\documentclass[a4paper,unpublished]{quantumarticle}
\pdfoutput=1

\usepackage{graphicx}
\graphicspath{{figures/}}
\usepackage{subcaption}
\usepackage{booktabs}
\usepackage{xcolor}
\usepackage{amsthm}

\usepackage[colorlinks, allcolors=quantumviolet]{hyperref}
\usepackage[numbers,sort&compress]{natbib}
\usepackage[
	print-unity-mantissa=false,
	range-phrase=--,
	group-minimum-digits=4,
]{siunitx}
\usepackage{algorithm}
\usepackage{algpseudocodex}
\usepackage{zref-clever}
\usepackage{tikz}

\usepackage{mathcommand}
\LoopCommands{\lettersAll}{\declaremathcommand#2{\textnormal{#1}}}

\newtheorem{lem}{Lemma}
\theoremstyle{definition}
\newtheorem{defn}{Definition}

\theoremstyle{remark}
\newtheorem{rem}{Remark}

\zcsetup{
    cap=true,
    countertype={defn=definition},
}

\usetikzlibrary{positioning, arrows.meta}

\newcommand\qmaddress{Quantum Motion,
9 Sterling Way,
London N7 9HJ,
United Kingdom}
\newcommand\oxfordaddress{Department of Materials,
University of Oxford,
Parks Road,
Oxford OX1 3PH,
United Kingdom}
\newcommand\osakaaddress{Center for Quantum Information and Quantum Biology,
The University of Osaka,
1-2 Machikaneyama,
Toyonaka 560-0043,
Japan}

\begin{document}
\title{Stream Decoding with Confidence Scores at Room and Cryogenic Temperatures}

\author{Maries Tahaab}
\email{maries@quantummotion.tech}
\affiliation{\qmaddress}

\author{Tim Chan}
\email{timothy.chan@materials.ox.ac.uk}
\affiliation{\oxfordaddress}
\affiliation{\osakaaddress}
\orcid{0000-0001-6187-7402}

\author{James Kirkman}
\email{james.kirkman@quantummotion.tech}
\affiliation{\qmaddress}

\author{Simon C. Benjamin}
\affiliation{\oxfordaddress}
\affiliation{\qmaddress}
\orcid{0000-0002-7766-5348}

\begin{abstract}
    In fault-tolerant quantum computing, fast and accurate decoding is crucial. Snowflake~\cite{chan2026snowflake} is
    a decoder for the surface code
    that runs in a streaming fashion. In this paper, we implement Snowflake on commercial FPGAs and validate them at room and cryogenic temperatures. Our results demonstrate high decoding throughput for small code distances that,
    when extrapolated,
    remains within acceptable limits for larger distances.
    Further, we incorporate the calculation of certain decoder confidence scores with negligible overhead both in terms of latency and physical resource utilisation.
	We note that implementing a large-scale system would require either a large FPGA beyond today's technology or clusters of FPGAs connected via a high-speed bus. Thus, we discuss an alternative architecture that exploits the locality of Snowflake by processing 2D slices of the 3D decoding window and offloading segments of the 3D structure to a high-speed memory.
\end{abstract}

\maketitle
\section{Introduction}\label{sec:introduction}

Fault-tolerant quantum computing involves representing quantum information robustly as logical qubits using a larger number of physical qubits~\cite{Shor1995,preskill1998fault}.
The relationship between the physical and the logical levels is defined by the chosen quantum error-correcting code (QECC). The physical qubits associated with a logical block have certain properties (generally, stabilisers i.e.\ parities of subsets of the physical qubits) that change only when errors occur. These stabilisers should therefore be measured frequently, and the classical information from each complete set of monitoring measurements is called the syndrome~\cite{terhal2015}.
This paper is relevant to the much-studied surface code:
a fast and hardware-friendly code that however requires many physical qubits to realise its logical qubits~\cite{Fowler2012a,Dennis2002,Litinski2019}.

A key task of the classical control hierarchy is to process these syndromes over the course of the quantum computation, and infer the likely nature of the errors that gave rise to them. The system responsible for this is the decoder; typically multiple decoders will be employed to monitor the many logical blocks in a utility-scale quantum computer~\cite{Maurya2026}. The rate at which a decoder can process syndrome information is crucial -- it must keep up with the rate at which syndromes are generated~\cite{Battistel2023};
this is called the \emph{inverse throughput}.
For fast solid state technologies such as superconducting, spin qubit, or photonic devices:
inverse throughput at or below \qty{1}{\micro s} per stabiliser measurement round is required~\cite[Figure~1]{Google2023} \cite[p~924]{Google2025}. For relatively slower atomic systems, the millisecond timescale may suffice~\cite{Bluvstein2026}.

Every logical qubit will ultimately be measured, either at the end of a computation to know the output or (more commonly) as a part of an internal step of the quantum algorithm such as gate teleportation~\cite{terhal2015,Fowler2013b}. Such a measurement involves measuring each physical qubit in a suitable basis, and presenting this data to the decoder which (from the monitoring of syndrome data to this point) is capable of interpreting the physical measurements as a given logical outcome: 0 or 1. The time needed to reach this decision generally differs from the inverse throughput and is longer; it is referred to as the measurement time or the \emph{reaction time}~\cite[Appendix~B.1]{gidney_2021}. The timescale is important since typically a computation cannot proceed (at least, not far) until this logical outcome is known. Failure to assign the correct outcome is a logical error in the computation; the decoder's capability to do this correctly is its \emph{accuracy}. The performance of a given decoder in this crucial task will depend on its sophistication and different decoders lead to logical error rates that vary by orders of magnitude even for relatively small codes~\cite[Figure~4]{senior2026aq2}. 

A decoder therefore has at least three core figures of merit:
its inverse throughput
(ideally sub-microsecond),
its reaction time
(preferably a few microseconds~\cite{gidney_2021,gidney2025factor2048bitrsa}),
and its accuracy; to this one can add practical considerations such as size and dollar cost. Unsurprisingly, these merits are in tension: faster timescales generally lead to lower accuracies.

While impressive progress has been made to achieve decoders that perform well by all metrics~\cite{senior2026aq2,grbic2026acceleratingtesseractdecoderquantum}, ideas have emerged to resolve the tension by having a two-tier decoder mechanism and employing {\it decoder switching}~\cite{toshio2025decoderswitching}. The first tier would be fast but not optimally accurate, while the second would be slower but more accurate.
For this to work, we would require a further feature for the tier-1 decoder:
a `soft output' that we call the \emph{decoder confidence score} (DCS).
This number should indicate how likely it is that {\it no logical error} would result
if the tier-1 decoder were to process its present syndrome
without recourse to the superior capabilities of tier-2.
When this confidence is high, there is no need for the tier-1 decoder to call for help.
This then introduces another figure of merit: the reliability of the DCS.
There are multiple possible ways by which a decoder can arrive at such a score,
with more reliable scores taking longer to compute.
The DCS is a feature that has recently been extensively researched~\cite{Smith2024,Bombin2024,meister2024efficientsoftoutputdecoderssurface,Gidney2025,lee2026efficientpostselectiongeneralquantum,Zhou2025b,dinca2026swimdistance,Birchall2026,Chen2025a,Dentelski2026,Mishima2026,Staples2026}.
In this work we implement and evaluate three DCSs
with extremely low computational overhead.

The majority of decoders that have been described in the literature are overlapping-window-based~\cite[\S VI.B]{Dennis2002}: they take in a block comprising multiple rounds of syndrome data (typically $d$ rounds where $d$ is the code distance) and process them collectively, during which time new rounds of syndrome data emerge from the QC and are stored until they constitute another block. This process imposes a coarse granularity to the computation, such that (for example) a logical measurement cannot be completed until relevant blocks are reconciled. An alternative and arguably superior mode of operation for a decoder is \emph{round-wise streaming}, i.e.\ taking in each round of syndrome data as it occurs and immediately adding it to a perpetually ongoing continuous decoding process. The decoder implemented and developed upon in this work, Snowflake~\cite{chan2026snowflake}, is a round-wise streaming decoder.

A final desirable feature of a decoder relates to its operating temperature. Many forms of quantum processing units need to be operated at \unit{\milli\kelvin} temperatures inside dilution refrigerators \cite{krantz2019, vandersypen2017, vandijk2019}. 
The need to convey measurement data out of the cryogenic system to room-temperature classical processors is a drawback in such systems. 
The required cabling introduces significant heat loads, and serialisation and deserialisation add latency; therefore, 
mature quantum computers would ideally process syndrome data within the cryogenic system~\cite{vandijk2019,Battistel2023}.
In this case, far less information,
such as the final result of logical measurement,
needs to be taken to room temperature. Operating in a cryogenic environment places severe restrictions on the heat generation of decoders -- at least the tier-1 decoders in the two-tier paradigm -- as we explore presently in this work.

Cryogenic QECC decoding has been previously studied in~\cite{ueno2021qecool,ueno2022qulatis,holmes2020nisqp}.
These decoders primarily utilise application-specific integrated circuits (ASICs)
based on single flux quantum and cryo-CMOS technologies.
Pinball~\cite{knapen2026pinball} takes a different approach by integrating a cryo-CMOS-based predecoder at the \qty{4}{\kelvin} stage of the cryostat to resolve sparse errors locally.
Its successor, Arqade~\cite{knapen2026mitigatingclassicalresourcecosts},
generates custom predecoders for quantum low-density parity-check codes.
Both significantly reduce the syndrome data bandwidth to the primary decoder.
While Arqade offers a field-programmable gate array (FPGA) implementation primarily for room-temperature operation, its cryogenic implementation remains ASIC-based.

Although ASICs offer exceptional power efficiency, they lack post-fabrication flexibility. 
In contrast, FPGAs allow for rapid modifications to the decoder, avoiding the long fabrication cycles associated with custom ASIC designs. 
These fabrication cycles can become a significant bottleneck given the rapid advancements in the field and the frequent need to update decoder logic. 
Given the recent surge in research focused on FPGA-based decoders
\cite{yang2026realtimesurfacecodeerrorcorrection,yan2026rethinkroleneuraldecoders,báscones2026scalablefpgaarchitecturerealtime,wegmann2026zerogpredecoderawaredecoderquantum,liyanage2024helios,Ziad2025,valentino2025quekuf,Wu2025_quantum_bibstyle}
and the need for rapid iteration, developing a fast and reliable method to evaluate FPGA designs directly at cryogenic temperatures is highly beneficial.

In \zcref{sec:background} we provide minimal preliminaries and review Snowflake,
then describe our hardware implementation in \zcref{sec:design_imp}.
We show FPGA utilisation
and decoding throughput results in \zcref{sec:results},
provided both by experiments and simulation.
\zcref[S]{sec:arch_2d} discusses the alternative 2D architecture
of Snowflake and its motivations.
\zcref[S]{sec:conclusion} concludes.

\section{Background}\label{sec:background}
Decoding the surface code (and many other QECCs) can be treated as a graph problem,
in which a subset of nodes,
termed \emph{defects},
must be paired either to each other or to one of the so-called
\emph{boundary nodes} of the \emph{decoding graph}.
The accuracy of the decoder is entirely dependent upon
these pairings,
and a reasonably accurate heuristic is
to pair together defects that are close to one another in the graph.
The Union--Find decoder (UF)~\cite{Delfosse2020,Delfosse2021}
approximates this heuristic
by growing exploratory regions,
called \emph{active clusters},
around each defect
and merging clusters that touch each other,
until they reach a stopping condition
(\emph{inactivation}),
at which point the pairings are
resolved within each cluster individually.

\begin{figure}
    \centering
    \begin{subfigure}[t]{\linewidth}
        \centering
        \includegraphics[width=0.72\linewidth]{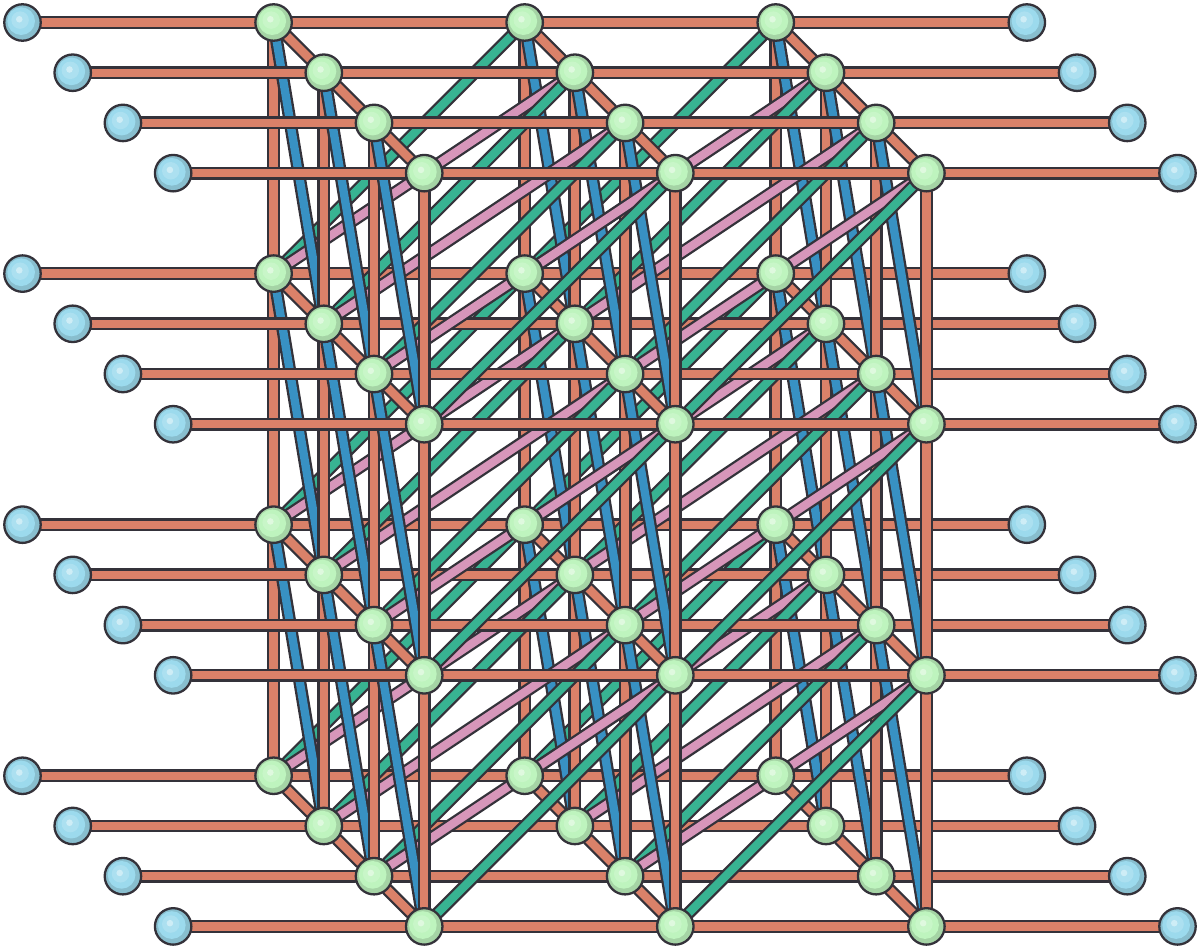}
        \caption{The decoding window $W$ used in Snowflake
		which is a subgraph of the full decoding graph
		that extends above and below $W$.
		Boundary nodes are coloured blue; otherwise, green.
		Nodes higher up in $W$ correspond to
		more recent stabiliser measurement data.}
        \label{fig:decoding_graph}
    \end{subfigure}

    \smallskip

    \begin{subfigure}[t]{\linewidth}
        \centering
        \includegraphics[width=0.72\linewidth]{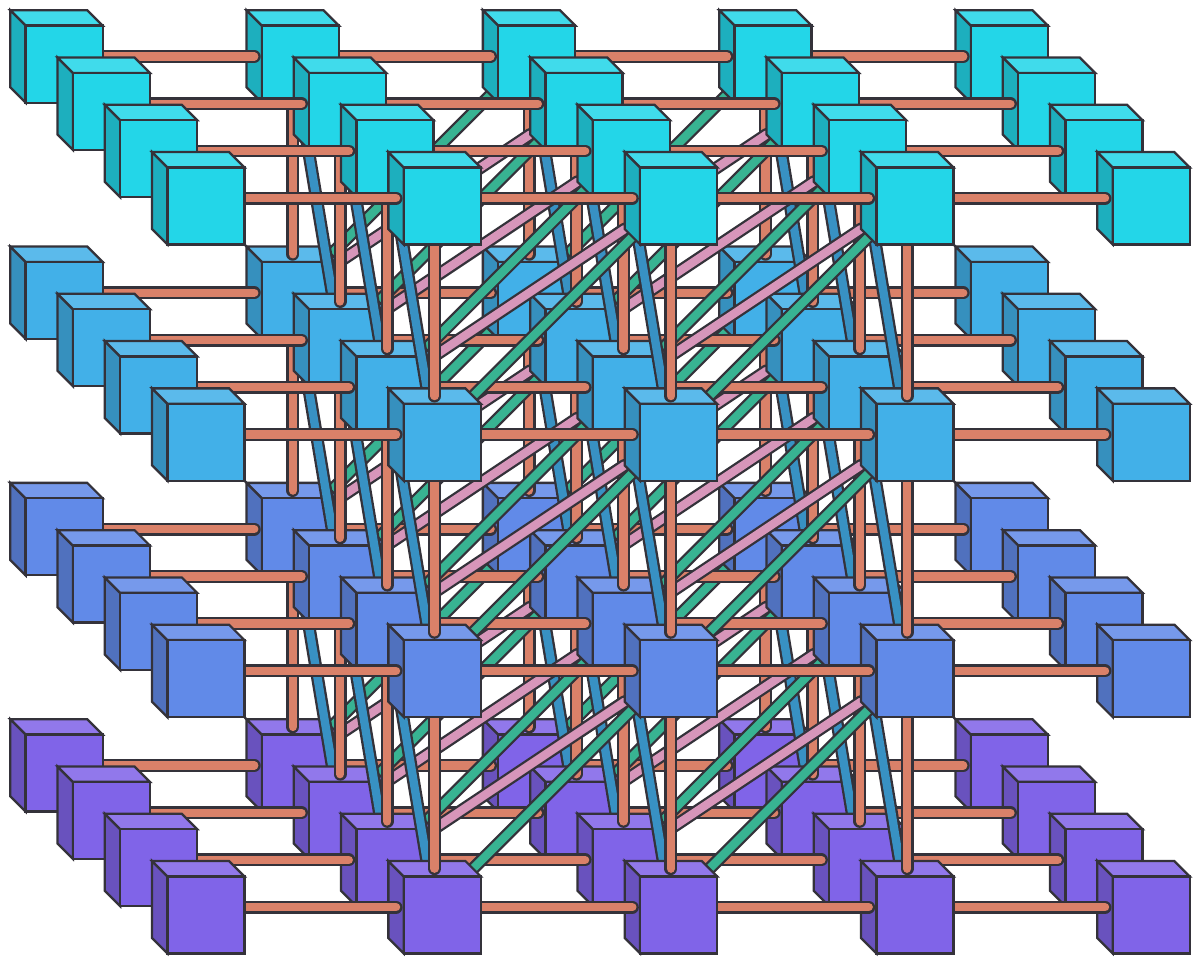}
        \caption{The distributed hardware implementation of Snowflake.
		Each cube represents a processing element corresponding to one node in $W$,
		and each tube represents a communication link corresponding to one edge.
		The centralised controller is not shown.
		This contrasts the 2D implementation proposed later in this work;
		cf.~\zcref{fig:2d_architecture}.}
        \label{fig:3d_distributed_implementation}
    \end{subfigure}

    \caption{Schematics relating to Snowflake configured for
	the distance-4 unrotated surface code.}
\end{figure}

Snowflake~\cite{chan2026snowflake} is based on UF,
but adapts this cluster-growing idea to the streaming case,
where the decoding graph is arbitrarily tall.
Snowflake considers a small connected subgraph of fixed size,
called the \emph{decoding window},
of the decoding graph
at any given time.
On the unrotated surface code,
the decoding window is a graph of $d^2(d+1)$ nodes
that is local when embedded in 3D,
as \zcref{fig:decoding_graph} illustrates.
Snowflake simultaneously
    slides this decoding window along the decoding graph,
    grows clusters gradually,
    and pairs defects within clusters.
A \emph{drop} is the process in which the window
slides up the decoding graph by one layer
(equivalent to a stabiliser measurement round),
so called because data appears to `drop' by one layer
from the perspective of the decoding window.

\begin{figure}
    \centering
    \includegraphics[width=0.45\textwidth]{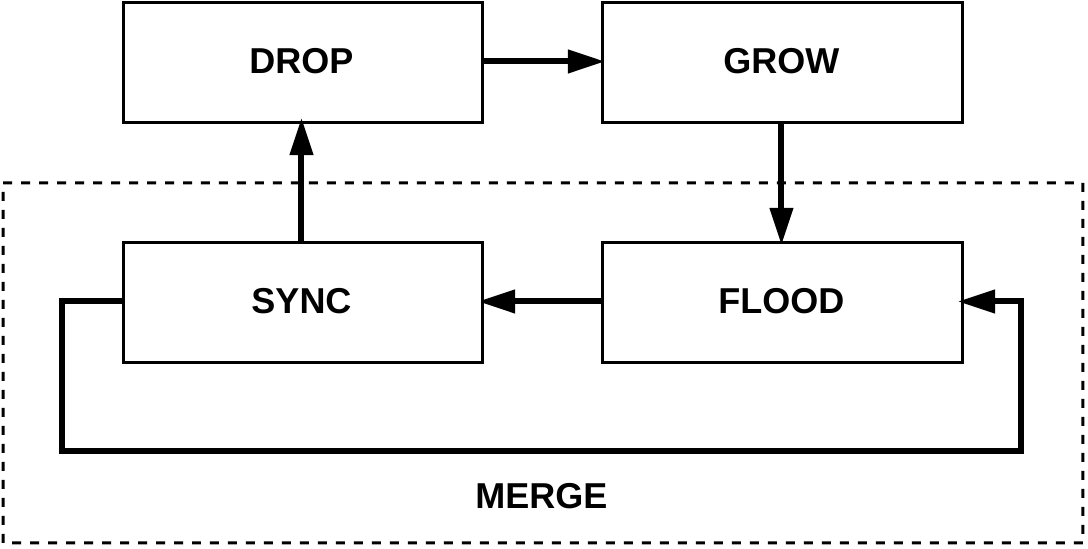}
    \caption{Simplified state transition diagram for Snowflake
	with the 1:1 cluster growth schedule.
	The analogous state diagram for the 2:1 schedule
	is shown in \zcref{fig:stage_flowchart}.}
    \label{fig:state_diagram_node}
\end{figure}

There are two versions of Snowflake,
defined by their so-called \emph{cluster growth schedule}:
the 1:1 and the 2:1 schedule.
The 2:1 schedule is more accurate,
but in this first work we
implement the 1:1 schedule for its simplicity,
leaving the 2:1 schedule for future work.
Snowflake with the 1:1 schedule operates by continuously repeating
a \emph{drop-grow-merge} cycle (see \zcref{fig:state_diagram_node}),
once per measurement round.
For interested readers,
a more detailed overview of Snowflake can be obtained
by reading \cite{chan2026snowflake} up to and including \S4.

\section{Design and Implementation}\label{sec:design_imp}

Since the decoder must have a sub-microsecond inverse throughput for superconducting quantum devices,
the metres-long cabling running from room-temperature electronics to the dilution refrigerator introduces a significant propagation delay. Furthermore, serialisation of the syndrome bitstream becomes inevitable, as routing a dedicated cable for each syndrome bit would conduct excessive heat, overwhelming the refrigerator's limited cooling capacity and degrading qubit coherence. 
Combined, the physical propagation delay and the overhead of serialising and deserialising the syndrome bitstream risk creating an error backlog \cite{terhal2015}. 
Minimising latency between the readout electronics and the decoder
is highly beneficial for overall system performance. To achieve this, we explored operating the decoder at cryogenic temperatures. This not only reduces communication latency but also lowers the power consumption of the digital transceivers feeding the syndrome bitstream
to the decoder, due to reduced transmitter drive power. 

This section describes the hardware implementation of the Snowflake decoder, along with the experimental setup developed to evaluate it at both room and cryogenic temperatures.
First, \zcref[S]{subsec:architecture} provides a high-level architectural overview of the hardware decoder, and \zcref[S]{subsec:dcs} explains the additional logic implemented to determine three decoder confidence scores.
Next, \zcref[S]{subsec:hardware} details the specific FPGAs selected for this application and the custom power delivery circuitry required to enable operation at cryogenic temperatures. 
\zcref[S]{subsec:setup} then describes the physical measurement setup and instrumentation used to characterise the decoder across both thermal environments. 
Finally, \zcref[S]{subsec:system_overview} provides a system-level overview and discusses key design choices for operating the decoder. 

\subsection{Architecture Overview}\label{subsec:architecture}
Our architecture of Snowflake consists of
a network of processing elements (PEs) and a centralised controller.
Each PE corresponds to a node and each communication link corresponds to an edge in the decoding window,
as \zcref{fig:3d_distributed_implementation} shows.
The controller synchronises the decoding process by issuing command signals to the PEs within the network.

The node is implemented as a finite state machine (FSM) in Verilog.
A simplified state diagram of this FSM is shown in \zcref{fig:state_diagram_node}.
Each node tracks its own state, and in turn the overall decoding process,
by maintaining a record of its variables within the FSM.
The variables belonging to each edge, however, are stored in a global register array implemented using the FPGA flip-flops, where each index uniquely maps to a specific edge within the network. Together, the PEs and the edge lookup table constitute the complete decoding graph.

For the clusters to grow---leading to defect annihilation and subsequently,
the generation of corrections---each node shares its variables with its immediate neighbours.
Consequently, it receives the variables being shared by those neighbours. By evaluating this shared data, the node FSM's logic governs the state transitions, thereby enabling a fully parallel decoding process.

\subsection{Decoder Confidence Scores}\label{subsec:dcs}
In addition to the core decoding logic,
we implement three hardware-efficient DCSs
with minimal impact on the resource utilisation of the FPGA,
as well as no time overhead
(achieved by computing the DCS
between consecutive drop cycles of the decoder).
The most accurate DCS we implement is called the
\textbf{Cluster Size 1-Norm Fraction}.
This metric (calculated by
summing a variable called \texttt{growth}
in the edge lookup table)
quantifies the total volume of the decoding window
occupied by any cluster.
Its utility has been explored in
\cite{lee2026efficientpostselectiongeneralquantum}:
for a distance-13 rotated surface code
under physical error rate \num{5e-3},
aborting the worst 30\% of shots according to this DCS
reduces the logical error rate by over an order of magnitude.

In \zcref{sec:other_decoder_confidence_scores}
we discuss our hardware implementation
of the two other lightweight DCSs,
although we observe them to be
less accurate than the cluster size 1-norm fraction.

\subsection{FPGA Selection and Power Delivery}\label{subsec:hardware}
The operation of commercial FPGAs in cryogenic environments has been previously demonstrated in
\cite{homulle2017cryoartix, jin2022cryokintex, lewis2025cryoartix, conwaylamb2016cryoartix}.
For our implementation, we chose the \textbf{AMD Kintex UltraScale+ XCKU5P} FPGA, fabricated on the \qty{16}{\nano\meter} FinFET+ process node
for its ample logic resources and strong potential for cryogenic operation.
Jin et al.~\cite{jin2022cryokintex} evaluated the AMD Kintex UltraScale+ XCKU3P FPGA, which has fewer logic resources but is also a member of the Kintex UltraScale+ family and is manufactured using the same process node.
In their evaluation, the XCKU3P demonstrated an $\sim$88\% reduction in static core current draw and a 30\% reduction in lookup table (LUT) delay compared to room-temperature operation.

We also evaluated the \textbf{AMD Artix-7 XC7A35T} FPGA, fabricated on the \qty{28}{\nano\meter} high performance-low power process node, due to prior successful demonstrations of AMD 7-series FPGAs operating at cryogenic temperatures~\cite{homulle2017cryoartix, lewis2025cryoartix, conwaylamb2016cryoartix}.
We selected this device for its reduced logic footprint, which results in lower power consumption in cryogenic environments compared to the much larger XCKU5P.

Delivering power to an FPGA inside a cryostat presents a challenge due to the unavoidable long interconnects and the step current load that occurs during device configuration. To negate the effects of line impedance from room-temperature power supplies and prevent brownouts, it was essential to place localised voltage regulators in close physical proximity to the FPGA. We therefore adapted a cryogenic low-dropout regulator based on the design in \cite{homulle2018cryoldo}, which ensured stable power delivery despite the long cabling.

\begin{figure}[htbp]
    \centering
    
    \begin{subfigure}[t]{0.45\textwidth}
		\centering
        \includegraphics[width=0.9\textwidth]{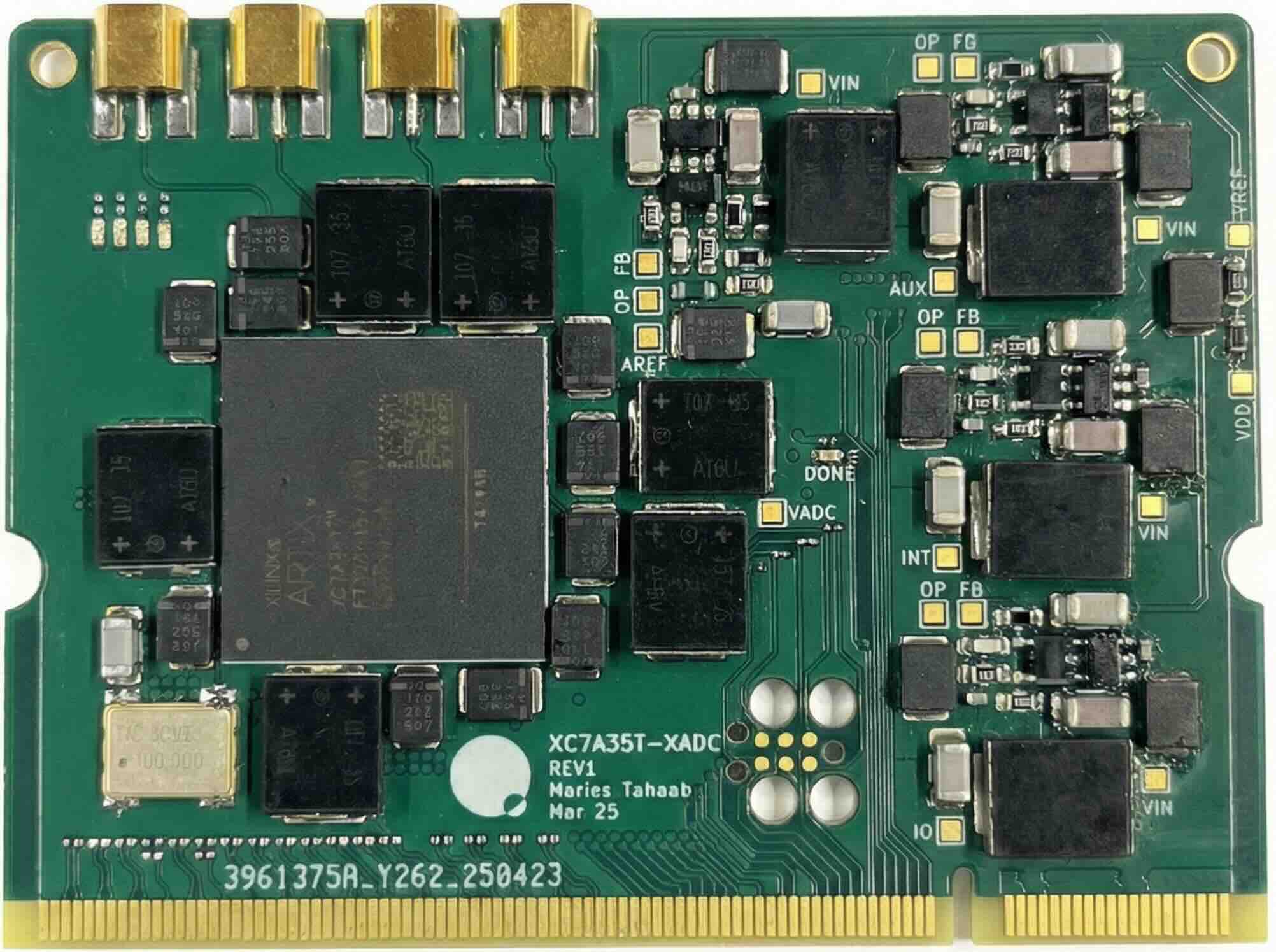}
		\caption{Artix-7 (XC7A35T-2FTG256I).}
    \end{subfigure}
    
	\smallskip

    \begin{subfigure}[t]{0.45\textwidth}
		\centering
        \includegraphics[width=0.9\textwidth]{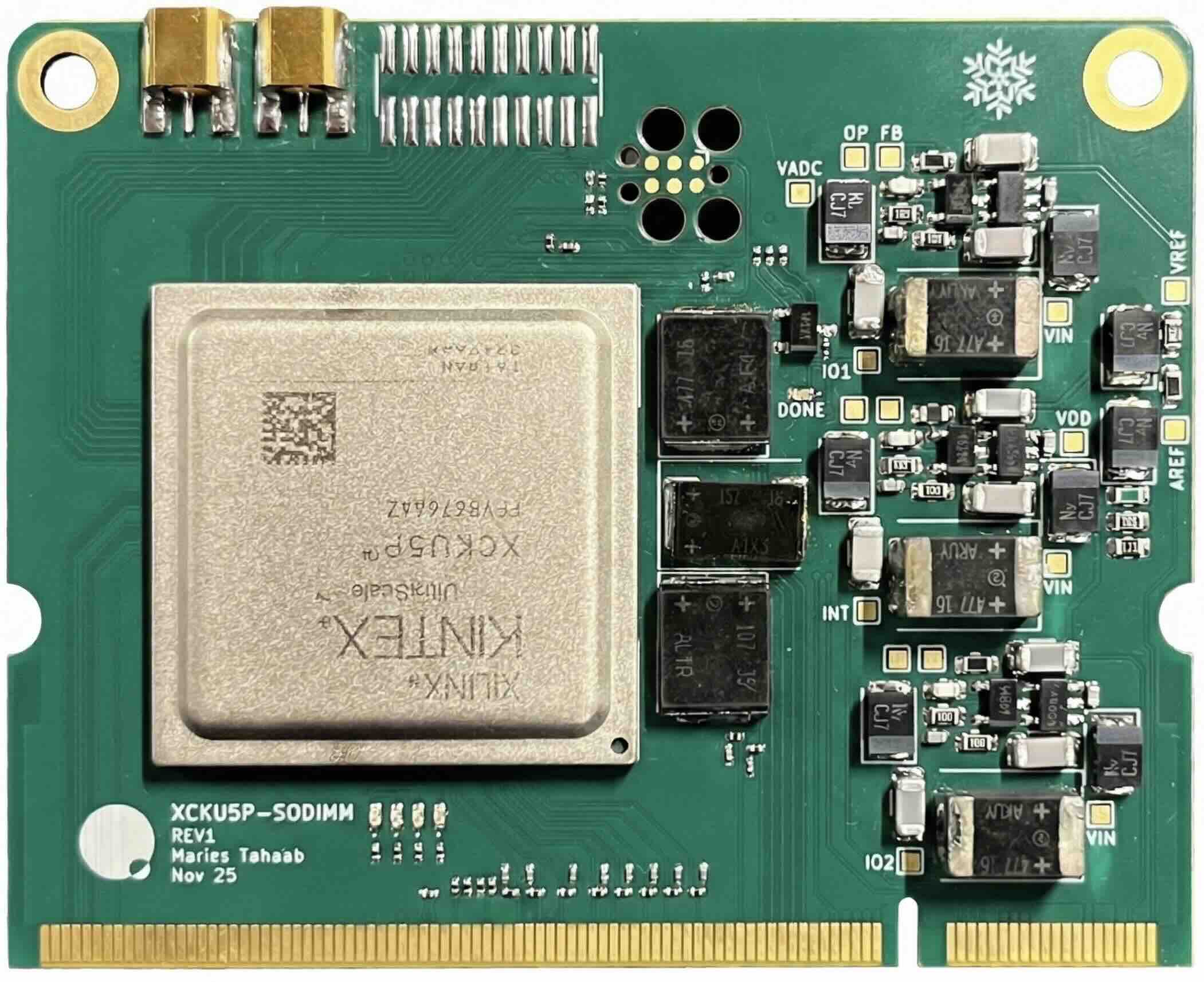}
        \caption{Kintex UltraScale+ (XCKU5P-2FFVP676I).}
    \end{subfigure}
    
    \caption{Custom printed circuit boards featuring
	onboard cryogenic voltage regulation for two FPGAs.}
    \label{fig:custom_pcbs}
\end{figure}

Custom printed circuit boards shown in \zcref{fig:custom_pcbs} were designed for both the Artix-7 and Kintex UltraScale+ FPGAs with onboard voltage regulation to test the decoder at cryogenic temperatures.

\subsection{Measurement Setup}\label{subsec:setup}
The room-temperature and cryogenic measurements for the FPGA modules were conducted in a front-loading cryostat equipped with a Sumitomo Heavy Industries RP-082B2 pulse tube cryocooler, which provides a cooling power of \qty{1}{\watt} at \qty{4}{\kelvin}. Room-temperature measurements were conducted with the cryocooler turned off.
\zcref[S]{fig:cryostat_setup} shows both the Kintex and Artix FPGA modules from \zcref{subsec:hardware} mounted inside the cryostat for testing.
\begin{figure}[htbp]
    \centering
    \includegraphics[width=0.45\textwidth]{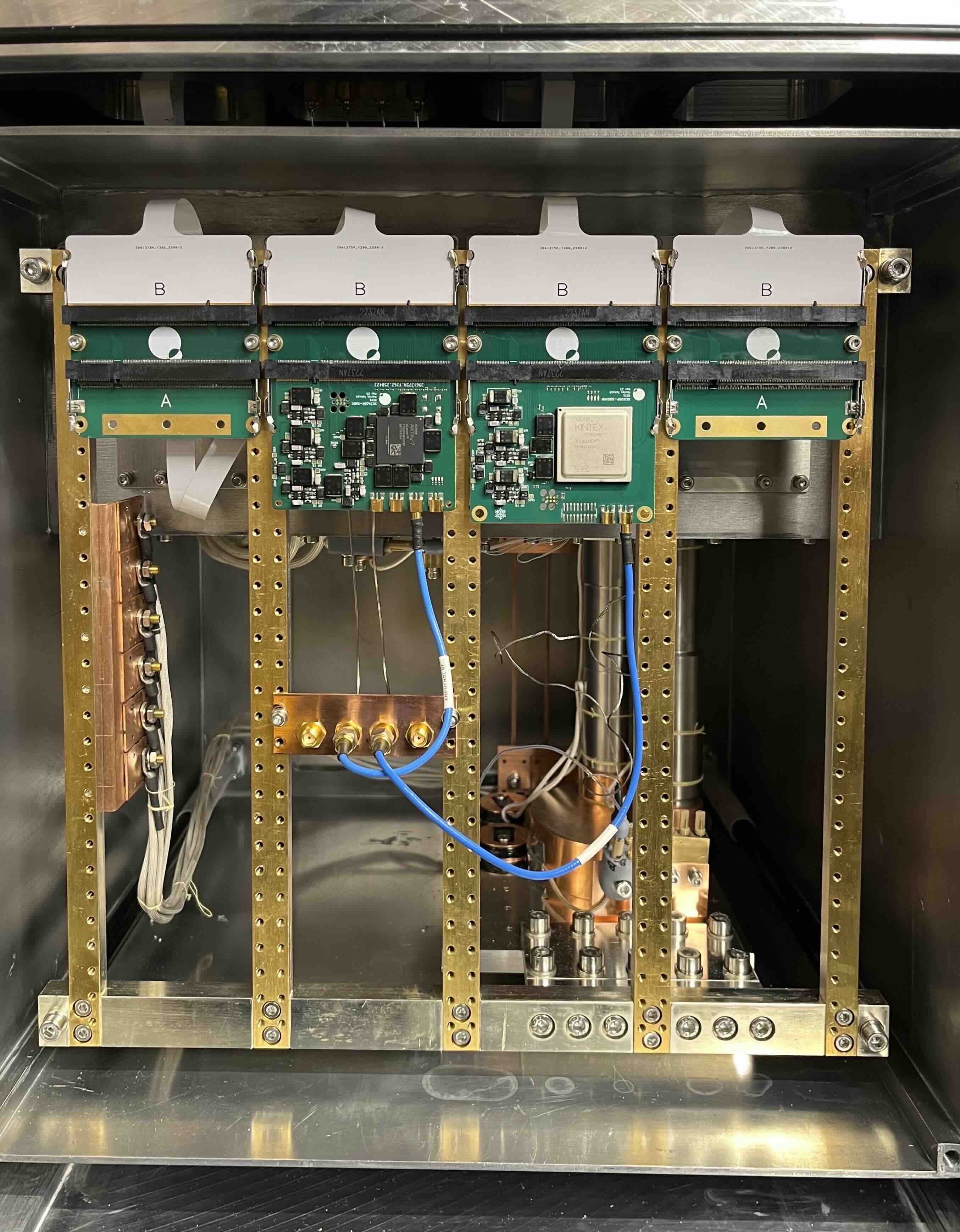}
    \caption{Measurement setup featuring both FPGA modules mounted to the cold plate of the front-loading cryostat.}
    \label{fig:cryostat_setup}
\end{figure}

Room-temperature instruments were used to drive and monitor the test setup. A Rigol DP811 linear DC power supply and a Yokogawa GS200 source measure unit (SMU) supplied power to the FPGAs. Additionally, a Rigol DG1032Z arbitrary waveform generator provided a \qty{20}{\mega\hertz} external reference clock. This clock was subsequently multiplied to the required \qty{200}{\mega\hertz} operating frequency for the decoder using the FPGA's internal Mixed-Mode Clock Manager.

To accurately determine the active die temperature during decoder operation, the internal temperature diodes of each FPGA were first calibrated against a Lake Shore Cryotronics Cernox CX-1030-HT sensor with the FPGAs turned off. A Keithley 2450 SMU was used to provide a constant \qty{10}{\micro\ampere} bias current to the diode while simultaneously measuring the forward voltage. A calibration curve from \qtyrange{4}{300}{\kelvin} was thus obtained, mapping the diode voltage to the temperature recorded by the Cernox sensor. This curve was then used to determine the active die temperature from the diode voltage while the decoder was in operation.

\subsection{System-Level Overview}\label{subsec:system_overview}
For testing and validation at cryogenic temperatures, the syndrome bitstream was transmitted to the cryogenic decoder from a host device at room temperature via a serial communication link. This same host was subsequently used to receive and validate the generated corrections.
A system-level block diagram of the complete setup is presented in \zcref[S]{fig:block_diagram_system}.
\begin{figure}[htbp]
    \centering
    \includegraphics[width=0.45\textwidth]{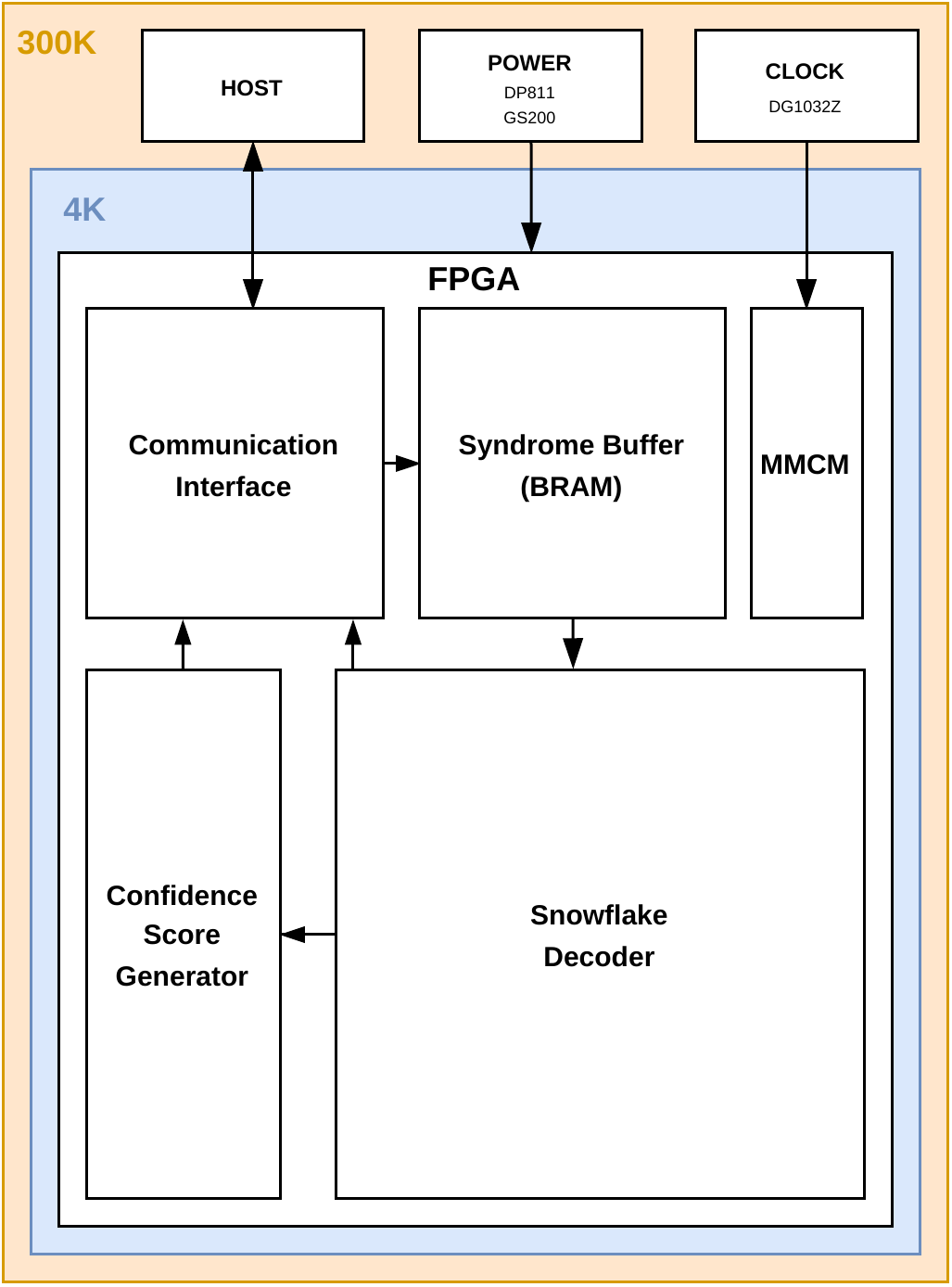}
    \caption{System-level block diagram for the decoder operating inside the cryostat.}
    \label{fig:block_diagram_system}
\end{figure}

To prevent the serial transmission protocol and the long cabling required by the cryostat from acting as a bottleneck to the decoder's operation, a syndrome buffer was implemented on the FPGA using its on-chip Block RAM (BRAM). The room-temperature host pre-loaded all generated syndromes into this buffer. Since the BRAM read latency is only a few clock cycles, it effectively eliminates the data transfer bottleneck. During operation, the BRAM loaded syndrome bitstreams into the decoder at every drop cycle, accurately simulating data reception from the local cryogenic electronics of a quantum device. Conversely, attempting to stream syndrome data from outside the cryostat during active decoding would introduce significant delays, making it difficult to accurately evaluate the decoder's true performance.

\section{Results}\label{sec:results}
This section presents the results obtained from evaluating the decoder across a range of code distances. The design was simulated for code distances $d=3$ to $21$ at a target clock frequency of \qty{200}{\mega\hertz}. To validate the simulated results, we present the results of physical hardware tests for distances $d = 3$ to $9$, utilising the experimental setup previously described in \zcref{subsec:setup}.

\subsection{FPGA Resource Utilisation}\label{subsec:res_utilisation}
\zcref[S]{tab:resource_utilisation} presents the resource utilisation of Snowflake
configured for circuit-level noise, across various code distances.
The data indicates a pronounced increase in FPGA resource utilisation as the code distance increases.
This scaling is a result of the number $d^2(d+1)$ of syndrome graph nodes,
and because the hardware architecture instantiates each node as an independent PE.
\begin{table}[htbp]
    \centering
    \caption{Resource utilisation for Snowflake across varying code distances.
	LUT = lookup table, FF = flip-flop, LUTRAM = lookup-table memory.
	$^*$Area-optimised design with reduced parallelism.}
    \label{tab:resource_utilisation}
    \begin{tabular}{@{} l r r r @{}}
        \toprule
        \textbf{Distance ($d$)} & \textbf{LUTs} & \textbf{FFs} & \textbf{LUTRAM} \\
        \midrule
        3 & \num{4 553} & \num{6 631} & \num{59} \\
        5 & \num{31 203} & \num{40 058} & \num{202} \\
        7 & \num{110 036} & \num{125 989} & \num{441} \\
        9 & \num{237 929} & \num{296 749} & \num{826} \\
        9$^*$ & \num{198 313} & \num{116 608} & -- \\
        \bottomrule
    \end{tabular}
\end{table}

For the $d=9$ implementation, the design exceeded the available LUT capacity of the target XCKU5P FPGA. To successfully fit the design onto the device, it was necessary to implement an area-optimised version of the decoder, which was achieved by reducing parallelisation and minimising pipeline stages within the PE logic. While these modifications allowed the decoder to fit within the hardware resources, the resulting increase in critical path length lowered the maximum operating clock frequency to \qty{60}{\mega\hertz}.

Additionally, this area-optimised approach resulted in a significant reduction in overall power consumption. Operating a highly utilised FPGA at high clock frequencies draws substantial current, which poses a physical risk of overloading the sensitive cryostat cabling. Consequently, even if the architecture could sustain a higher operating frequency, the clock rate would have required throttling to remain within the safe current-carrying limits of the experimental setup. A detailed analysis of the decoder's power consumption and thermal characteristics is presented in \zcref{subsec:res_cryo_performance}.

\subsection{Decoder Throughput}\label{subsec:res_throughput}
In this section, we discuss the timing performance of Snowflake in terms of the average decoding time across varying code distances.
The decoder was evaluated on hardware for $d=3$ to $9$ and in simulation for distances up to $d=21$ at a target clock frequency of \qty{200}{\mega\hertz}.
Syndromes were generated by simulating 1024 stabiliser measurement rounds
under a circuit-level noise model
(the same model as in \cite[\S B]{Chan2023c})
of physical error rate $p = \num{1e-3}$:
roughly seven times below threshold.
Decoding time was computed by
multiplying the clock period by the number of clock cycles.
\zcref[S]{fig:decoder_throughput_comparison} reports
the mean decoding time for $d=3$ to $21$.
\begin{figure}[htbp]
    \centering
    \includegraphics[width=1.0\columnwidth]{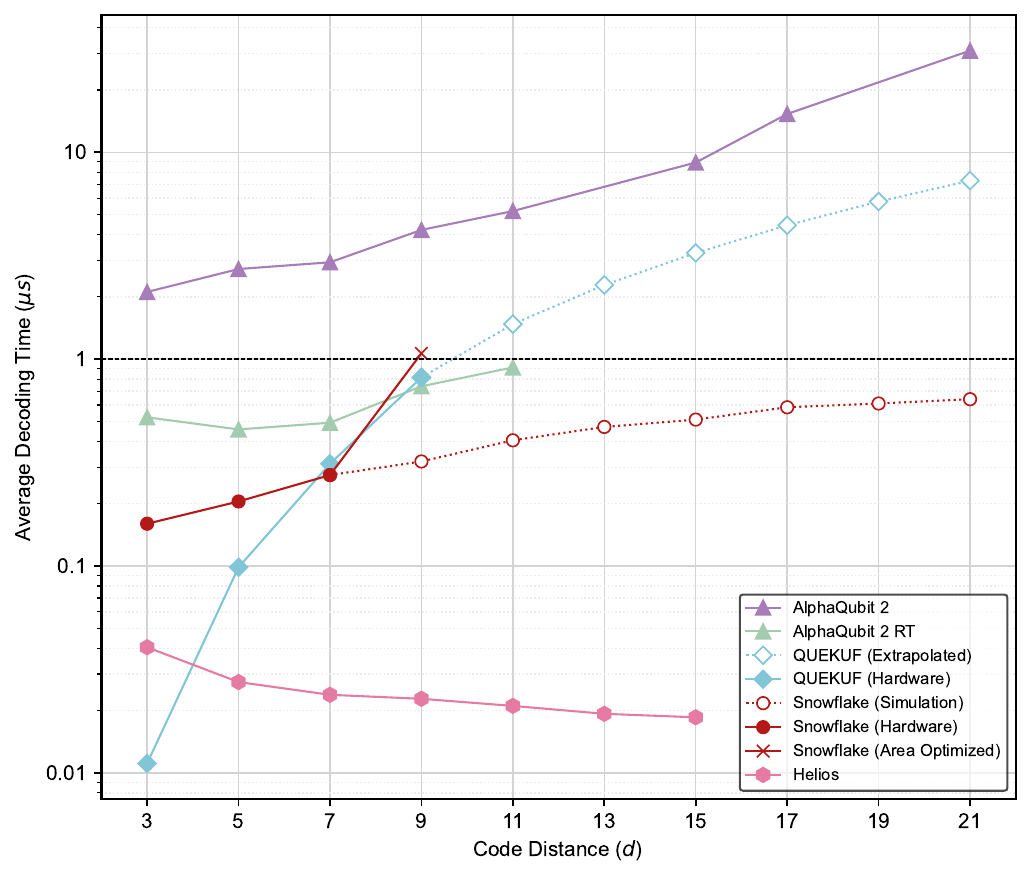}
    \caption{Mean duration to decode per stabiliser measurement round,
    plotted against code distance.
    The curve for Snowflake (simulation)
    agrees well with its software analogue
    \cite[Figure~9d green curve]{chan2026snowflake}.}
    \label{fig:decoder_throughput_comparison}
\end{figure}

\begin{table*}[htbp]
\centering
\caption{Summary of hardware decoder implementations.
UF = Union--Find,
TPU = tensor processing unit,
ML = machine learning,
SC = surface code,
CC = colour code.}
\label{tab:decoder_metrics_summary}
\begin{tabular}{@{}llllc@{}}
\toprule
\textbf{Decoder} & \textbf{Hardware Device} & \textbf{Algorithm/Approach} & \textbf{Targeted Code} & \textbf{Max Distance ($d$)} \\ \midrule
Helios           & AMD VCU129               & UF                 & Rotated SC        & 51                 \\
Snowflake        & AMD XCKU5P               & UF                 & Unrotated SC      & 9                  \\
QUEKUF           & AMD U55C                 & UF                 & Toric Code        & 10                 \\
AQ2              & Google Trillium TPU      & ML                 & Rotated SC \& CC  & 23 (SC), 27 (CC)   \\
AQ2-RT           & Google Trillium TPU      & ML                 & Rotated SC \& CC  & 23 (SC), 27 (CC)   \\ \bottomrule
\end{tabular}
\end{table*}

Snowflake achieves a sub-microsecond average decoding time for code distances up to $d=21$.
This satisfies the \qty{1}{\micro\second} time budget typical of superconducting quantum devices, making it a viable candidate for real-time decoding in superconducting platforms, as well as in platforms with more relaxed timing constraints, such as electron spin and neutral atom qubits.
The deviation from the simulated data observed at $d=9$ is a result of the area-optimised implementation
running at a reduced clock frequency.
Furthermore, the architecture leaves ample room for further timing optimisation. Increasing the clock frequency beyond \qty{200}{\mega\hertz} would result in a considerable reduction in decoding time, albeit at the cost of increased power consumption.

We first compare our implementation with AlphaQubit 2 (AQ2) and its compact, real-time variant, AQ2-RT~\cite{senior2026aq2}.
These are machine learning-based decoders evaluated on the rotated surface code and colour code.
As shown in \zcref[S]{fig:decoder_throughput_comparison}, the average decoding time of AQ2 remains well above \qty{1}{\micro\second} across all evaluated code distances.
On the other hand, AQ2-RT achieves sub-microsecond decoding times up to $d=11$, though this comes at the cost of a slight increase in logical error rate compared to AQ2.
The decoding time of Snowflake is, on average, approximately \qty{350}{\nano\second} lower than that of AQ2-RT.
It is important to note that both AQ2 and AQ2-RT were evaluated at a physical error rate of $p = \num{1.5e-3}$, which is higher than the error rate of \num{1e-3} used in our evaluation.
As a result, the difference in throughput would likely be reduced if AQ2-RT were evaluated at an error rate similar to ours.

We then consider QUEKUF~\cite{valentino2025quekuf}, another UF-based hardware decoder evaluated on the toric code. 
Since real-time hardware data for QUEKUF is only available up to $d=10$, we estimate its decoding speed for larger distances using the extrapolated clock-cycle counts provided in their study.
For this comparison, we use the clock cycle count data at an error rate of $p = \num{1e-3}$ under a code-capacity noise model.
We optimistically assume a constant clock frequency of \qty{247.5}{\mega\hertz}, which corresponds to their reported maximum operating frequency at $d = 10$.
Based on these figures, QUEKUF's average decoding time remains under \qty{1}{\micro\second} for distances ranging from $d=3$ to $9$.
However, it crosses the \qty{1}{\micro\second} threshold for $d \ge 11$, making it suitable for real-time decoding in superconducting based quantum devices only up to $d=9$.

Finally, we compare our implementation with Helios \cite{liyanage2024helios},
another UF-based hardware decoder that uses the overlapping-window approach to stream decoding.
Helios was evaluated on the rotated surface code under a circuit-level noise model at $p = \num{1e-3}$. As shown in \zcref[S]{fig:decoder_throughput_comparison},
its average decoding time per stabiliser measurement round decreases with increasing code distance $d$. As a result, it achieves sub-microsecond decoding times for large code distances, demonstrating the lowest decoding time of any hardware decoder reported in the literature to date.

\zcref[S]{tab:decoder_metrics_summary} provides a summary of the hardware platforms, algorithms, targeted QECC, and maximum evaluated code distances for each decoder discussed in this section.

\subsection{Performance at Cryogenic Temperatures}\label{subsec:res_cryo_performance}
The power consumption and thermal characteristics of the decoder implemented on the XCKU5P FPGA across various code distances are depicted in \zcref[S]{fig:decoder_power_and_thermals}. Power consumption is evaluated at both room ($P_\textnormal{rt}$) and cryogenic temperatures ($P_\textnormal{cryo}$), along with the corresponding FPGA die ($T_\textnormal{die}$) and cryostat ($T_\textnormal{rail}$) temperatures.
\begin{figure}[htbp]
    \centering
    \includegraphics[width=1.0\columnwidth]{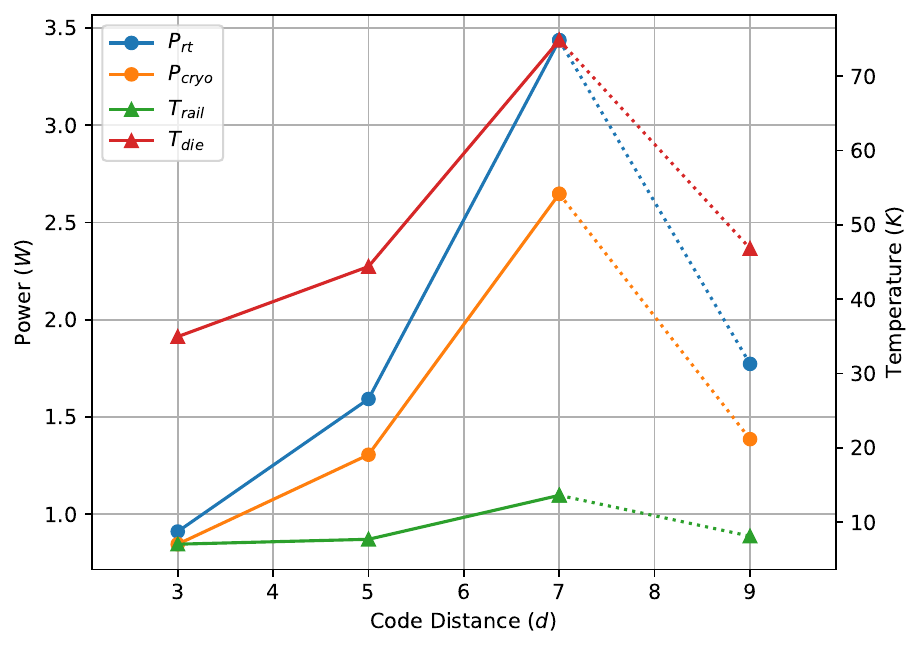}
    \caption{Impact of code distance ($d$) on decoder power consumption at room ($P_\textnormal{rt}$) and cryogenic ($P_\textnormal{cryo}$) temperatures, along with the corresponding XCKU5P FPGA die ($T_\textnormal{die}$) and cryostat ($T_\textnormal{rail}$) temperatures.}
    \label{fig:decoder_power_and_thermals}
\end{figure}

The findings indicate that operating the FPGA in a cryogenic environment consistently results in lower power consumption compared to room temperature.
While this difference reaches a maximum of $\sim$\qty{0.8}{\watt} for $d=7$, the overall reduction was not as substantial as demonstrated in \cite{jin2022cryokintex}.

For smaller code distances ($d=3, 5$), the cryostat rail temperature remains between \qty{7}{\kelvin} and \qty{8}{\kelvin} despite the increasing power draw.
On the contrary, the highly utilised $d=7$ implementation causes a sharp increase,
drawing \qty{2.65}{\watt} under cryogenic conditions and driving $T_\textnormal{rail}$ up to \qty{13.62}{\kelvin}.
To mitigate extreme power dissipation for the $d=9$ implementation, the reduced clock speed of the area-optimised architecture successfully lowered the power draw (denoted by the dotted lines) to \qty{1.39}{\watt} and $T_\textnormal{rail}$ to \qty{8.13}{\kelvin}, albeit with an increase in the average decoding time from \num{0.32} to \qty{1.06}{\micro\second} per measurement round.

Due to its limited resource capacity compared to the XCKU5P, the XC7A35T FPGA could only accommodate the $d=3$ decoder operated at \qty{200}{\mega\hertz}. Decoder implementations for $d > 3$ exceeded the available resources on this device. 
During cryogenic operation of the $d=3$ decoder on the XC7A35T FPGA, $T_\textnormal{die}$ ($T_\textnormal{rail}$) was measured to be \qty{38.2}{\kelvin} (\qty{5.5}{\kelvin}), 
with a power draw $P_\textnormal{cryo}$ ($P_\textnormal{rt}$) of \qty{0.448}{\watt} (\qty{0.351}{\watt}).
This amounts to a $\sim$47\% reduction in power draw at cryogenic temperatures compared to the XCKU5P FPGA.

Furthermore, the cryostat temperature consistently exceeded \qty{4}{\kelvin} throughout the active operation of the decoders. While this continuous heat dissipation did not cause operational disruptions in our front-loading cryostat, deploying such a system near the lower stages of a dilution refrigerator would introduce a substantial heat load. This thermal load risks raising the base temperature of the cryostat, which would severely affect overall system performance.

\section{An Alternative 2D Architecture}\label{sec:arch_2d}
\begin{figure}[h!]
    \centering
    \includegraphics[width=0.8\linewidth]{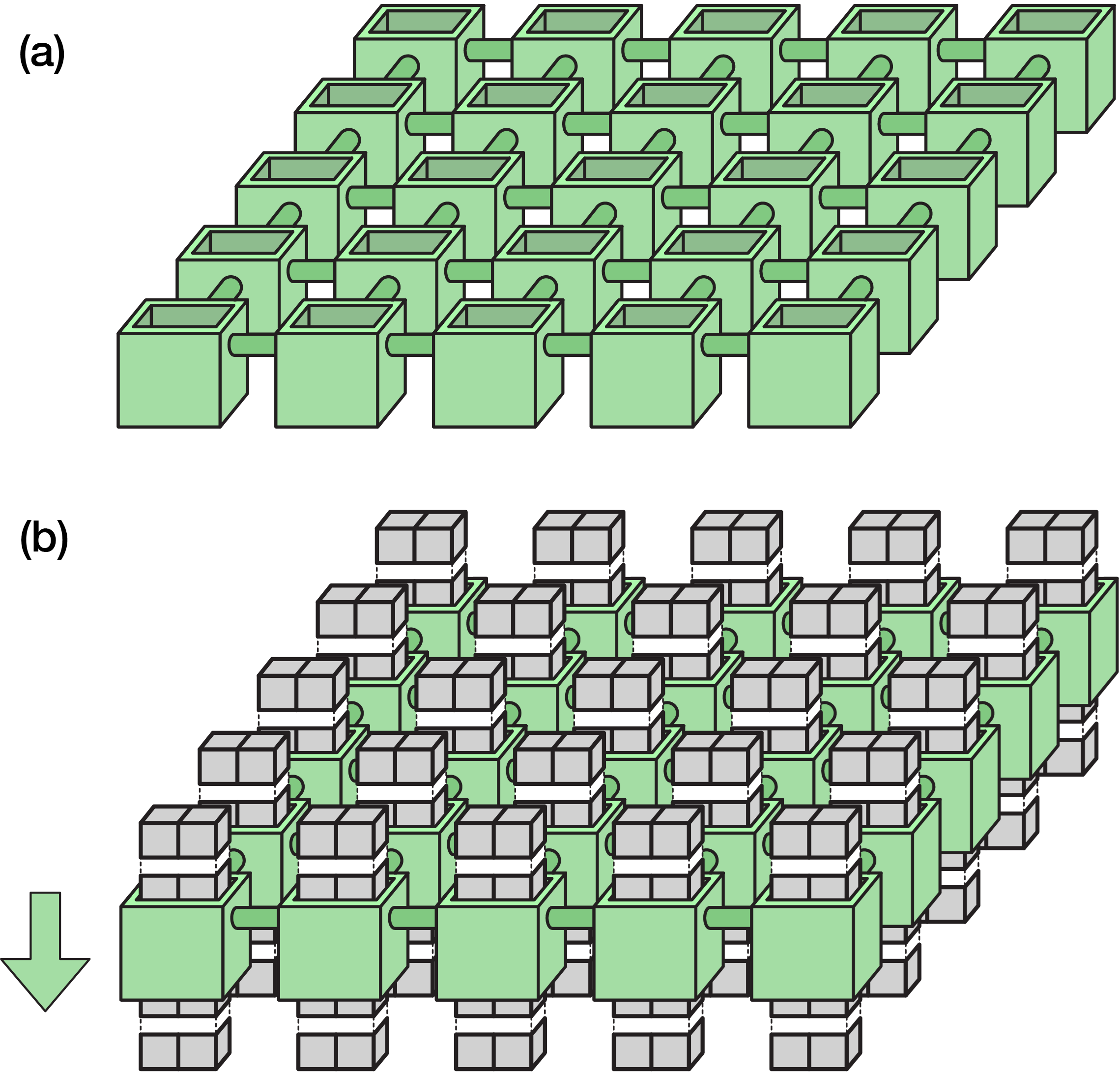}
    \caption{(a) A flattened implementation of Snowflake
    in which each processing element corresponds to a vertical column of nodes in $W$.
    (b) The 2D grid of processing elements in this implementation
	processes the decoding window one horizontal sheet at a time.
	This approach contrasts with the idea of implementing all $d^2(d+1)$ nodes,
	as in the earlier \zcref{fig:3d_distributed_implementation}.}
    \label{fig:2d_architecture}
\end{figure}

In this section we propose an alternative architecture for Snowflake:
instead of processing all nodes in the decoding window simultaneously,
process them one sheet at a time from top to bottom,
as \zcref{fig:2d_architecture} shows.

\zcref[S]{subsec:res_active_depth} motivates this architecture
with the observation that most of the PEs in our existing architecture are underutilised.
\zcref[S]{sec:main_idea} discusses the hardware setup of such an architecture,
and \zcref{sec:early_stop} proposes an early-stop criterion
to achieve even higher decoding throughput.

\subsection{Motivating Observation: Active Depth}\label{subsec:res_active_depth}
To evaluate the extent to which the 3D decoding network of \zcref{sec:design_imp} is utilised,
we first define a node as \emph{active} iff it is part of an active cluster.
We also associate each node in the decoding window with a \emph{depth};
nodes at the top (bottom) of the window have depth 1 ($h$),
where $h$ is the height of the window.
We then analyse the following.
\begin{defn}\label{defn:active_depth}
The \emph{active depth} is the depth of the deepest active node
in the decoding window.
If no node is active,
the active depth is zero.
\end{defn}
\noindent
This determines the deepest sheet at which nontrivial merging
(the stage that dominates the runtime of Snowflake)
occurs.
A small active depth indicates that most of the decoding window
contains no active clusters and is thus idle.
A large active depth indicates there is an active node
near the bottom of the window,
and hence near-complete utilisation of the decoding network.

\zcref[S]{fig:active_depth} shows the distribution of active depth
for code distances ranging from $d=3$ to $25$
under the same noise model used in \zcref{subsec:res_throughput}.
It is clear that the decoding activity is predominantly confined
to the upper sheets of the decoding network:
most of the PEs
(corresponding to nodes in the lower region of the decoding window)
are underutilised,
as they hardly ever participate in non-trivial merging.
This motivates the alternative architecture proposed in this section.

\begin{figure}[htbp]
    \centering
    \includegraphics[width=1.0\columnwidth]{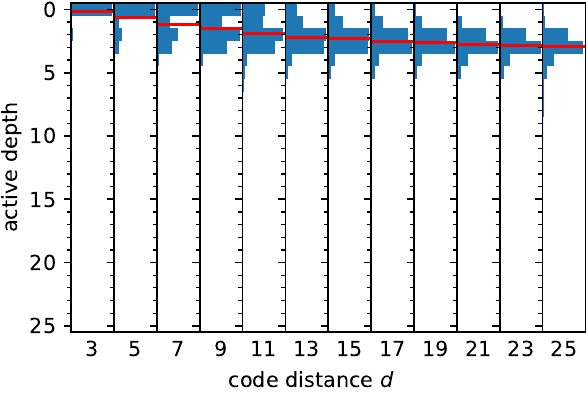}
    \caption{The distribution of \emph{active depth}
    (\zcref{defn:active_depth})
    in Snowflake with the 2:1 cluster growth schedule,
	under $p=10^{-3}$ circuit-level noise.
    Within each histogram,
    the red horizontal line shows the mean across $10^2 d$ stabiliser measurement rounds.
	\zcref[S]{sec:additional_active_depth_plot} shows that
	this mean scales logarithmically with code distance.
	Data obtained from emulation using a custom Python package~\cite{Chan2023a_quantum_bibstyle}.}
    \label{fig:active_depth}
\end{figure}

\subsection{Hardware Setup}\label{sec:main_idea}
\begin{figure}[htbp]
    \centering
    \includegraphics[width=0.9\columnwidth]{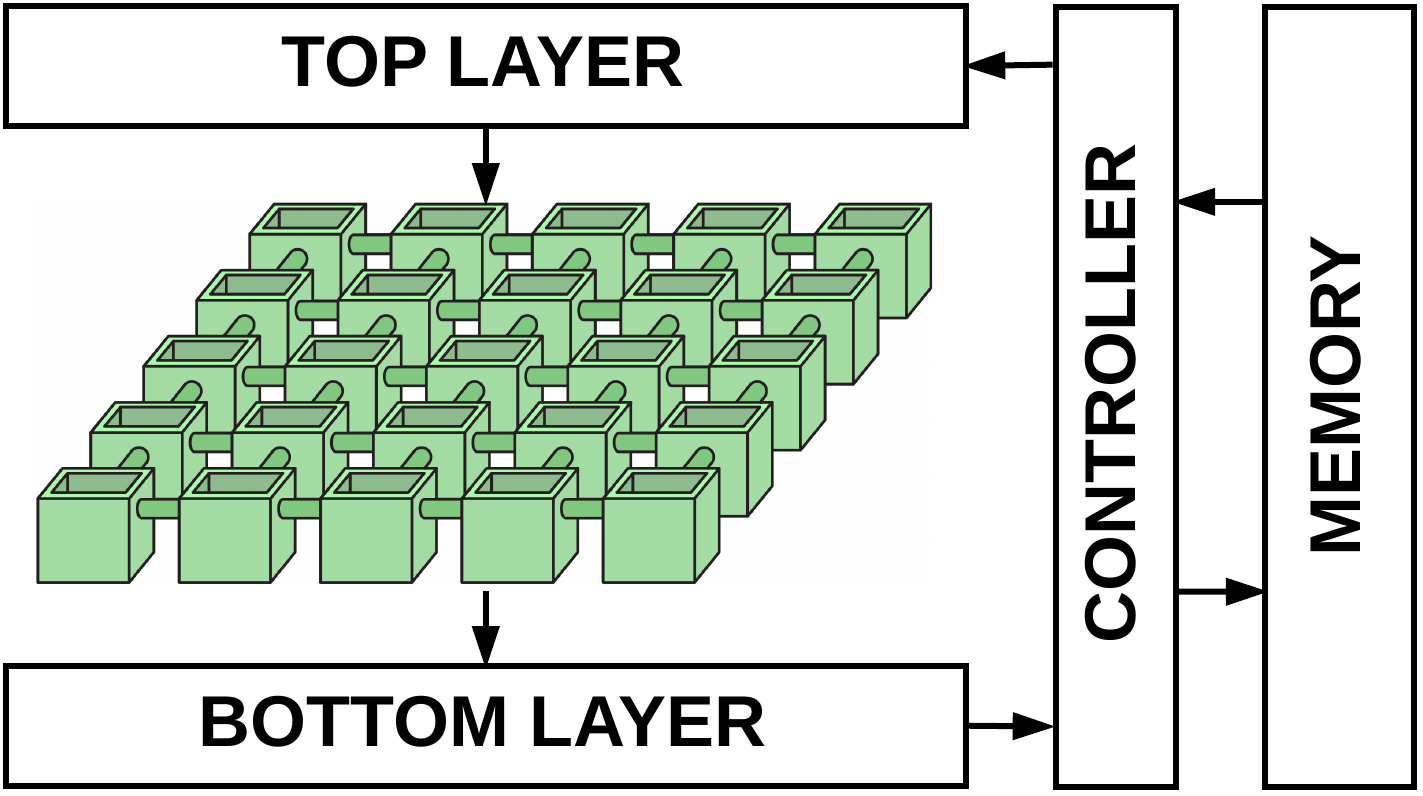}
    \caption{An architecture for the flattened implementation of Snowflake.
	A processing layer operates on the data contained within it, referencing the top and bottom layers held in local registers to advance the algorithm. The controller manages the data flow, loading distant layers of the 3D decoding network from memory into these reference registers as the algorithm progresses.}
    \label{fig:decoder_sliced}
\end{figure}

The alternative architecture would require only a 2D grid of processors
(that does not scale with the height of the decoding window),
along with high-speed memory,
as shown in \zcref{fig:decoder_sliced}.
The grid represents a horizontal 2D slice of the 3D decoding network,
while the node variables,
stored in the memory,
are fed to the 2D grid.
In this way,
the hardware preserves the local nature of, and exactly implements,
the Snowflake algorithm.

Storing decoder variables in an external memory
allows parallel logical blocks to access this data simultaneously.
This accessibility enables parallelised implementations
of various DCS generators
(such as for swim distance~\cite{dinca2026swimdistance,kishi2026efficientsoftoutputdecodingextracluster}),
and decoder switching to stronger CPU/GPU-based decoders \cite{toshio2025decoderswitching}
if the same memory is shared between the strong and weak decoders.
These features could be implemented with minimal time overhead,
albeit with a slight increase in overall resource utilisation.

Pseudocode for this alternative architecture is given in \zcref{sec:pseudocode_for_the_2d_architecture}.
We note similar time-multiplexing strategies to address hardware scalability
have also been discussed in \cite{kishi2026efficientsoftoutputdecodingextracluster}.
This modification would use substantially less FPGA logic (by a factor of $d$),
though the clock frequency would need to be increased to prevent backlog.

\subsection{Early-Stop Logic}\label{sec:early_stop}
Our analysis in \zcref{subsec:res_active_depth}
indicates that decoder activity rarely reaches deep into the decoding window,
with corrections predominantly completed in the top region.
The pseudocode in \zcref{sec:pseudocode_for_the_2d_architecture}
exploits this tendency
by skipping computation for sheets deeper than the active depth;
all the nodes in these sheets are inactive,
so we are guaranteed that such computation will not change their variables.
Snowflake is thus exactly implemented even with this early-stop logic,
while a measurable speed-up in decoding throughput is achieved.

\section{Conclusion}\label{sec:conclusion}
We have presented a cryogenic implementation of Snowflake,
equipped with confidence scores that introduce no timing overhead.
Our results indicate that operating the decoder in a cryogenic environment generates a substantial amount of heat due to its high FPGA resource utilisation. 
Consequently, deploying this implementation at the \qty{4}{\kelvin} stage in a large-scale system is impractical, 
as the dissipated heat would elevate the temperatures of the lower stages of the dilution refrigerator and degrade qubit fidelities. 
A potential approach for practical cryogenic decoding would involve deploying clusters of small FPGA-based decoders, 
connected via a high-speed bus \cite{liyanage2024multifpga, liyanage2025network}, 
into the higher-temperature stages (\qty{50}{\kelvin}) of the dilution refrigerator, 
where the available cooling power is significantly greater.

Alternatively, to mitigate these hardware constraints from an architectural perspective,
we have proposed a 2D architecture for Snowflake.
This design would drastically reduce the decoder's logic footprint
and subsequent power consumption with negligible impact on throughput.
The reduction in power consumption would enable
the hardware to operate safely within the strict thermal limits
of the \qty{4}{\kelvin} stage,
while serving as a foundation for a custom cryo-CMOS ASIC
optimised for low-temperature operation.
We believe this is a highly promising direction left for future investigation.

\begin{acknowledgments}
MT thanks Jonathan Warren and Alexander Waterworth for cryogenic technical support
and Holly Farler for useful discussions.
TC acknowledges support from
an EPSRC DTP studentship and
JST ASPIRE Japan Grant Number JPMJAP2319.
We acknowledge
the use of
the University of Oxford Advanced Research Computing
(ARC)
facility~\cite{Richards2015_quantum_bibstyle},
and three EPSRC projects:
QCS Hub (EP/T001062/1),
RoaRQ (EP/W032635/1),
and SEEQA (EP/Y004655/1).
\end{acknowledgments}

\section*{Competing Interests}
Two relevant patent applications exist:
WO2025238191A1 and EP26195274.1.

\bibliographystyle{quantum}
\bibliography{references}

@article{Battistel2023,
	title = {Real-time decoding for fault-tolerant quantum computing: progress, challenges and outlook},
	author = {Francesco Battistel and Christopher Chamberland and Kauser Johar and Ramon W. J. Overwater and Fabio Sebastiano and Luka Skoric and Yosuke Ueno and Muhammad Usman},
	year = 2023,
	month = aug,
	publisher = {IOP Publishing},
	journal = {Nano Futures},
	volume = {7},
	number = {3},
	pages = {032003},
	doi = {10.1088/2399-1984/aceba6},
}

@article{Bombin2024,
	title = {Fault-Tolerant Postselection for Low-Overhead Magic State Preparation},
	author = {Bomb\'{\i}n, H\'ector and Pant, Mihir and Roberts, Sam and Seetharam, Karthik I.},
	year = 2024,
	month = jan,
	publisher = {American Physical Society},
	journal = {PRX Quantum},
	volume = {5},
	issue = {1},
	pages = {010302},
	numpages = {19},
	doi = {10.1103/PRXQuantum.5.010302},
}

@article{chan2026snowflake,
	title = {{Snowflake}: A Distributed Streaming Decoder},
	author = {Chan, Tim},
	year = 2026,
	month = mar,
	publisher = {{Verein zur F{\"{o}}rderung des Open Access Publizierens in den Quantenwissenschaften}},
	journal = {{Quantum}},
	volume = {10},
	pages = {2033},
	issn = {2521-327X},
	doi = {10.22331/q-2026-03-20-2033},
}

@article{Chan2023c,
	title = {{Actis}: A Strictly Local {Union}--{Find} Decoder},
	author = {Chan, Tim and Benjamin, Simon C.},
	year = 2023,
	month = nov,
	publisher = {{Verein zur F{\"{o}}rderung des Open Access Publizierens in den Quantenwissenschaften}},
	journal = {{Quantum}},
	volume = {7},
	pages = {1183},
	issn = {2521-327X},
	doi = {10.22331/q-2023-11-14-1183},
}

@article{Fowler2012a,
	title = {Surface codes: Towards practical large-scale quantum computation},
	author = {Fowler, Austin G. and Mariantoni, Matteo and Martinis, John M. and Cleland, Andrew N.},
	year = 2012,
	month = sep,
	publisher = {American Physical Society},
	journal = {Physical Review A},
	volume = {86},
	issue = {3},
	pages = {032324},
	numpages = {48},
	doi = {10.1103/PhysRevA.86.032324},
}

@misc{meister2024efficientsoftoutputdecoderssurface,
      title={Efficient soft-output decoders for the surface code}, 
      author={Nadine Meister and Christopher A. Pattison and John Preskill},
      year={2024},
      eprint={2405.07433},
      archivePrefix={arXiv},
      primaryClass={quant-ph},
      url={https://arxiv.org/abs/2405.07433}, 
}

@article{lee2026efficientpostselectiongeneralquantum,
   title={Efficient post-selection for general quantum {LDPC} Codes},
   volume={12},
   ISSN={2056-6387},
   url={http://dx.doi.org/10.1038/s41534-026-01242-x},
   DOI={10.1038/s41534-026-01242-x},
   number={1},
   journal={npj Quantum Information},
   publisher={Springer Science and Business Media LLC},
   author={Lee, Seok-Hyung and English, Lucas H. and Bartlett, Stephen D.},
   year={2026},
   month=Apr }

@INPROCEEDINGS{liyanage2024multifpga,
    author={Liyanage, Namitha and Wu, Yue and Houghton, Emmet and Zhong, Lin},
    booktitle={2024 IEEE International Conference on Quantum Computing and Engineering (QCE)}, 
    title={Multi-{FPGA} System for Quantum Error Correction with Lattice Surgery}, 
    year={2024},
    volume={02},
    number={},
    pages={622-623},
    doi={10.1109/QCE60285.2024.10435}
}

@article{homulle2018cryoldo,
    title = {Cryogenic low-dropout voltage regulators for stable low-temperature electronics},
    journal = {Cryogenics},
    volume = {95},
    pages = {11-17},
    year = {2018},
    issn = {0011-2275},
    doi = {https://doi.org/10.1016/j.cryogenics.2018.08.006},
    url = {https://www.sciencedirect.com/science/article/pii/S001122751830198X},
    author = {Harald Homulle and Edoardo Charbon}
}

@INPROCEEDINGS{homulle2017cryoartix,
    author={Homulle, Harald and Charbon, Edoardo},
    booktitle={2017 International Conference on Field Programmable Technology (ICFPT)}, 
    title={Performance characterization of {Altera} and {Xilinx} 28 nm {FPGAs} at cryogenic temperatures}, 
    year={2017},
    volume={},
    number={},
    pages={25-31},
    doi={10.1109/FPT.2017.8280117}
}

@article{conwaylamb2016cryoartix,
   title={An FPGA-based instrumentation platform for use at deep cryogenic temperatures},
   volume={87},
   ISSN={1089-7623},
   url={http://dx.doi.org/10.1063/1.4939094},
   DOI={10.1063/1.4939094},
   number={1},
   journal={Review of Scientific Instruments},
   publisher={AIP Publishing},
   author={Conway Lamb, I. D. and Colless, J. I. and Hornibrook, J. M. and Pauka, S. J. and Waddy, S. J. and Frechtling, M. K. and Reilly, D. J.},
   year={2016},
   month=Jan
}

@misc{kishi2026efficientsoftoutputdecodingextracluster,
    title={Even More Efficient Soft-Output Decoding with Extra-Cluster Growth and Early Stopping}, 
    author={Kaito Kishi and Riki Toshio and Jun Fujisaki and Hirotaka Oshima and Shintaro Sato and Keisuke Fujii},
    year={2026},
    eprint={2602.03336},
    archivePrefix={arXiv},
    primaryClass={quant-ph},
    url={https://arxiv.org/abs/2602.03336}, 
}

@misc{lewis2025cryoartix,
    title={Implementation of Field Programmable Gate Arrays ({FPGAs}) in Extremely Cold Environments for Space and Cryogenic Computing Applications}, 
    author={Christopher Lewis and Drew Sellers and Michael Hamilton},
    year={2025},
    eprint={2504.13305},
    archivePrefix={arXiv},
    primaryClass={eess.SY},
    url={https://arxiv.org/abs/2504.13305}
}

@INPROCEEDINGS{jin2022cryokintex,
    author={Jin, Zhanhong and Wang, Xinzhe and Liang, Futian and Peng, Cheng-Zhi},
    booktitle={2022 IEEE 65th International Midwest Symposium on Circuits and Systems (MWSCAS)}, 
    title={Cryogenic Characterization of Commercial Devices for Application of Quantum Computing Electronics}, 
    year={2022},
    volume={},
    number={},
    pages={1-5},
    doi={10.1109/MWSCAS54063.2022.9859526}
}

@misc{dinca2026swimdistance,
    title={Error mitigation for logical circuits using decoder confidence}, 
    author={Maria Dincă and Tim Chan and Simon C. Benjamin},
    year={2026},
    eprint={2512.15689},
    archivePrefix={arXiv},
    primaryClass={quant-ph},
    url={https://arxiv.org/abs/2512.15689}
}

@article{Smith2024,
	title = {Mitigating errors in logical qubits},
	author = {Smith, Samuel C. and Brown, Benjamin J. and Bartlett, Stephen D.},
	year = 2024,
	month = nov,
	journal = {Communications Physics},
	volume = {7},
	number = {1},
	pages = {386},
	issn = {2399-3650},
	doi = {10.1038/s42005-024-01883-4},
}

@misc{toshio2025decoderswitching,
    title={Decoder Switching: Breaking the Speed-Accuracy Tradeoff in Real-Time Quantum Error Correction}, 
    author={Riki Toshio and Kaito Kishi and Jun Fujisaki and Hirotaka Oshima and Shintaro Sato and Keisuke Fujii},
    year={2025},
    eprint={2510.25222},
    archivePrefix={arXiv},
    primaryClass={quant-ph},
    url={https://arxiv.org/abs/2510.25222}, 
}

@misc{Zhou2025b,
	title = {Error Mitigation of Fault-Tolerant Quantum Circuits with Soft Information},
	author = {Zeyuan Zhou and Shaun Pexton and Aleksander Kubica and Yongshan Ding},
	year = 2025,
	eprint = {2512.09863},
	archiveprefix = {arXiv},
	primaryclass = {quant-ph},
}

@article{terhal2015,
  title = {Quantum error correction for quantum memories},
  author = {Terhal, Barbara M.},
  journal = {Rev. Mod. Phys.},
  volume = {87},
  issue = {2},
  pages = {307--346},
  numpages = {40},
  year = {2015},
  month = {Apr},
  publisher = {American Physical Society},
  doi = {10.1103/RevModPhys.87.307},
  url = {https://link.aps.org/doi/10.1103/RevModPhys.87.307}
}

@manual{Richards2015_quantum_bibstyle,
	title = {\href{https://doi.org/10.5281/zenodo.22558}{University of Oxford Advanced Research Computing}},
	author = {Richards, Andrew},
	year = 2015,
	month = aug,
	doi = {10.5281/zenodo.22558},
}

@repository{Chan2023a_quantum_bibstyle,
	author = {Chan, Tim},
	year = 2023,
	url = {https://github.com/timchan0/localuf},
}

@article{Delfosse2020,
	title = {Linear-time maximum likelihood decoding of surface codes over the quantum erasure channel},
	author = {Delfosse, Nicolas and Z\'emor, Gilles},
	year = 2020,
	month = jul,
	publisher = {American Physical Society},
	journal = {Physical Review Research},
	volume = {2},
	issue = {3},
	pages = {033042},
	numpages = {5},
	doi = {10.1103/PhysRevResearch.2.033042},
}

@article{Delfosse2021,
	title = {Almost-linear time decoding algorithm for topological codes},
	author = {Delfosse, Nicolas and Nickerson, Naomi H.},
	year = 2021,
	month = dec,
	publisher = {{Verein zur F{\"{o}}rderung des Open Access Publizierens in den Quantenwissenschaften}},
	journal = {{Quantum}},
	volume = {5},
	pages = {595},
	issn = {2521-327X},
	doi = {10.22331/q-2021-12-02-595},
}

@article{Bluvstein2026,
	title = {A fault-tolerant neutral-atom architecture for universal quantum computation},
	author = {Bluvstein, Dolev and Geim, Alexandra A. and Li, Sophie H. and Evered, Simon J. and Bonilla Ataides, J. Pablo and Baranes, Gefen and Gu, Andi and Manovitz, Tom and Xu, Muqing and Kalinowski, Marcin and Majidy, Shayan and Kokail, Christian and Maskara, Nishad and Trapp, Elias C. and Stewart, Luke M. and Hollerith, Simon and Zhou, Hengyun and Gullans, Michael J. and Yelin, Susanne F. and Greiner, Markus and Vuletić, Vladan and Cain, Madelyn and Lukin, Mikhail D.},
	year = 2026,
	month = jan,
	publisher = {Nature Publishing Group UK London},
	journal = {Nature},
	volume = {649},
	number = {8095},
	pages = {39--46},
	issn = {1476-4687},
	doi = {10.1038/s41586-025-09848-5},
}

@article{preskill1998fault,
  title={Fault-tolerant quantum computation},
  author={Preskill, John},
  journal={Introduction to quantum computation and information},
  volume={213},
  year={1998},
  publisher={World Scientific},
  doi={10.1142/3724},
}

@article{Litinski2019,
	title = {A Game of Surface Codes: Large-Scale Quantum Computing with Lattice Surgery},
	author = {Litinski, Daniel},
	year = 2019,
	month = mar,
	publisher = {{Verein zur F{\"{o}}rderung des Open Access Publizierens in den Quantenwissenschaften}},
	journal = {{Quantum}},
	volume = {3},
	pages = {128},
	issn = {2521-327X},
	doi = {10.22331/q-2019-03-05-128},
}

@inproceedings{Maurya2026,
	title = {A Case for Elastic Quantum Error Correction Decoders},
	author = {Maurya, Satvik and Molavi, Abtin and Albarghouthi, Aws and Tannu, Swamit},
	year = 2026,
	publisher = {Association for Computing Machinery},
	address = {New York, NY, USA},
	isbn = {9798400722127},
	booktitle = {Proceedings of the 21st European Conference on Computer Systems},
	pages = {514--531},
	numpages = {18},
	doi = {10.1145/3767295.3803584},
	location = {McEwan Hall/The University of Edinburgh, Edinburgh, Scotland UK},
	series = {EUROSYS '26},
}

@misc{Fowler2013b,
	title = {Time-optimal quantum computation},
	author = {Austin G. Fowler},
	year = 2013,
	eprint = {1210.4626},
	archiveprefix = {arXiv},
	primaryclass = {quant-ph},
}

@article{Shor1995,
	title = {Scheme for reducing decoherence in quantum computer memory},
	author = {Shor, Peter W.},
	year = 1995,
	month = oct,
	publisher = {American Physical Society},
	journal = {Physical Review A},
	volume = {52},
	issue = {4},
	pages = {R2493--R2496},
	numpages = {0},
	doi = {10.1103/PhysRevA.52.R2493},
}

@misc{gidney2025factor2048bitrsa,
      title={How to factor 2048 bit {RSA} integers with less than a million noisy qubits},
      author={Craig Gidney},
      year={2025},
      eprint={2505.15917},
      archivePrefix={arXiv},
      primaryClass={quant-ph},
      url={https://arxiv.org/abs/2505.15917}, 
}

@article{Gidney2025,
	title = {Yoked surface codes},
	author = {Gidney, Craig and Newman, Michael and Brooks, Peter and Jones, Cody},
	year = 2025,
	month = may,
	journal = {Nature Communications},
	volume = {16},
	number = {1},
	pages = {4498},
	issn = {2041-1723},
	doi = {10.1038/s41467-025-59714-1},
}

@misc{grbic2026acceleratingtesseractdecoderquantum,
      title={Accelerating the Tesseract Decoder for Quantum Error Correction}, 
      author={Dragana Grbic and Laleh Aghababaie Beni and Noah Shutty},
      year={2026},
      eprint={2602.02985},
      archivePrefix={arXiv},
      primaryClass={quant-ph},
      url={https://arxiv.org/abs/2602.02985}, 
}

@article{Google2023,
	title = {Suppressing quantum errors by scaling a surface code logical qubit},
	author = {{Google Quantum AI}},
	year = 2023,
	month = feb,
	publisher = {Nature Publishing Group UK London},
	journal = {Nature},
	volume = {614},
	number = {7949},
	pages = {676--681},
	issn = {1476-4687},
	doi = {10.1038/s41586-022-05434-1},
}

@article{Google2025,
	title = {Quantum error correction below the surface code threshold},
	author = {{Google Quantum AI} and {collaborators}},
	year = 2025,
	month = feb,
	publisher = {Springer Science and Business Media LLC},
	journal = {Nature},
	volume = {638},
	number = {8052},
	pages = {920--926},
	issn = {1476-4687},
	doi = {10.1038/s41586-024-08449-y},
	url = {http://dx.doi.org/10.1038/s41586-024-08449-y},
}

@article{krantz2019,
   title={A quantum engineer’s guide to superconducting qubits},
   volume={6},
   ISSN={1931-9401},
   url={http://dx.doi.org/10.1063/1.5089550},
   DOI={10.1063/1.5089550},
   number={2},
   journal={Applied Physics Reviews},
   publisher={AIP Publishing},
   author={Krantz, P. and Kjaergaard, M. and Yan, F. and Orlando, T. P. and Gustavsson, S. and Oliver, W. D.},
   year={2019},
   month=jun,
}

@article{vandersypen2017,
	title={Interfacing spin qubits in quantum dots and donors—hot, dense, and coherent},
	volume={3},
	ISSN={2056-6387},
	url={http://dx.doi.org/10.1038/s41534-017-0038-y},
	DOI={10.1038/s41534-017-0038-y},
	number={1},
	journal={npj Quantum Information},
	publisher={Springer Science and Business Media LLC},
	author={Vandersypen, L. M. K. and Bluhm, H. and Clarke, J. S. and Dzurak, A. S. and Ishihara, R. and Morello, A. and Reilly, D. J. and Schreiber, L. R. and Veldhorst, M.},
	year={2017},
	month=sep,
}

@article{vandijk2019,
	title = {The electronic interface for quantum processors},
	journal = {Microprocessors and Microsystems},
	volume = {66},
	pages = {90-101},
	year = {2019},
	issn = {0141-9331},
	doi = {https://doi.org/10.1016/j.micpro.2019.02.004},
	url = {https://www.sciencedirect.com/science/article/pii/S0141933118303417},
	author = {J.P.G. {van Dijk} and E. Charbon and F. Sebastiano},
}

@inproceedings{holmes2020nisqp,
	author = {Holmes, Adam and Jokar, Mohammad Reza and Pasandi, Ghasem and Ding, Yongshan and Pedram, Massoud and Chong, Frederic T.},
	title = {{NISQ+}: boosting quantum computing power by approximating quantum error correction},
	year = {2020},
	isbn = {9781728146614},
	publisher = {IEEE Press},
	url = {https://doi.org/10.1109/ISCA45697.2020.00053},
	doi = {10.1109/ISCA45697.2020.00053},
	booktitle = {Proceedings of the ACM/IEEE 47th Annual International Symposium on Computer Architecture},
	pages = {556–569},
	numpages = {14},
	location = {Virtual Event},
	series = {ISCA '20}
}

@misc{Birchall2026,
	title = {{Macromux}: scalable postselection for high-threshold fault-tolerant quantum computation},
	author = {Patrick Birchall and Jacob Bridgeman and Christopher Dawson and Terry Farrelly and Yehua Liu and Naomi Nickerson and Mihir Pant and Sam Roberts and Karthik Seetharam and David Tuckett},
	year = 2026,
	eprint = {2603.04875},
	archiveprefix = {arXiv},
	primaryclass = {quant-ph},
}

@misc{Chen2025a,
	title = {Scalable accuracy gains from postselection in quantum error correcting codes},
	author = {Hongkun Chen and Daohong Xu and Grace M. Sommers and David A. Huse and Jeff D. Thompson and Sarang Gopalakrishnan},
	year = 2025,
	eprint = {2510.05222},
	archiveprefix = {arXiv},
	primaryclass = {cond-mat.stat-mech},
}

@misc{Dentelski2026,
	title = {Neural network decoder confidence as a learned proxy for the logical gap},
	author = {David Dentelski},
	year = {2026},
	eprint = {2606.08758},
	archiveprefix = {arXiv},
	primaryclass = {quant-ph},
}

@misc{Mishima2026,
	title = {Adaptive Window Decoding based on Spatiotemporal Complementary Gap},
	author = {Moeto Mishima and Riki Toshio and Kaito Kishi and Jun Fujisaki and Hirotaka Oshima and Shintaro Sato and Keisuke Fujii},
	year = 2026,
	eprint = {2605.14637},
	archiveprefix = {arXiv},
	primaryclass = {quant-ph},
}

@misc{Staples2026,
	title = {Scalable Postselection of Quantum Resources},
	author = {J. Wilson Staples and Winston Fu and Jeff D. Thompson},
	year = 2026,
	eprint = {2603.08697},
	archiveprefix = {arXiv},
	primaryclass = {quant-ph},
}

@misc{wegmann2026zerogpredecoderawaredecoderquantum,
      title={{Zero-G}: A Pre-Decoder-Aware Decoder for Quantum Error Correction}, 
      author={Peter Wegmann and Theofilos Augoustis and Aleksandra Świerkowska and Emmanouil Giortamis and Pramod Bhatotia},
      year={2026},
      eprint={2608.02030},
      archivePrefix={arXiv},
      primaryClass={quant-ph},
      url={https://arxiv.org/abs/2608.02030}, 
}

@article{Ziad2025,
	title = {Local clustering decoder as a fast and adaptive hardware decoder for the surface code},
	author = {Ziad, Abbas B. and Zalawadiya, Ankit and Topal, Canberk and Camps, Joan and Geh{\'e}r, Gy{\"o}rgy P. and Stafford, Matthew P. and Turner, Mark L.},
	year = 2025,
	month = dec,
	publisher = {Springer Science and Business Media LLC},
	journal = {Nature Communications},
	volume = {16},
	number = {1},
	pages = {11048},
	issn = {2041-1723},
	doi = {10.1038/s41467-025-66773-x},
}

@inproceedings{Wu2025_quantum_bibstyle,
	title = {{Micro} {Blossom}: Accelerated Minimum-Weight Perfect Matching Decoding for Quantum Error Correction},
	author = {Wu, Yue and Liyanage, Namitha and Zhong, Lin},
	year = 2025,
	publisher = {Association for Computing Machinery},
	address = {New York, NY, USA},
	isbn = {9798400710797},
	booktitle = {Proceedings of the 30th ACM International Conference on Architectural Support for Programming Languages and Operating Systems, Volume 2},
	pages = {639--654},
	numpages = {16},
	doi = {10.1145/3676641.3716005},
	location = {Rotterdam, Netherlands},
}

@INPROCEEDINGS{ueno2021qecool,
  author={Ueno, Yosuke and Kondo, Masaaki and Tanaka, Masamitsu and Suzuki, Yasunari and Tabuchi, Yutaka},
  booktitle={2021 58th ACM/IEEE Design Automation Conference (DAC)}, 
  title={{QECOOL}: On-Line Quantum Error Correction with a Superconducting Decoder for Surface Code}, 
  year={2021},
  volume={},
  number={},
  pages={451-456},
  doi={10.1109/DAC18074.2021.9586326}
}

@INPROCEEDINGS{ueno2022qulatis,
  author={Ueno, Yosuke and Kondo, Masaaki and Tanaka, Masamitsu and Suzuki, Yasunari and Tabuchi, Yutaka},
  booktitle={2022 IEEE International Symposium on High-Performance Computer Architecture (HPCA)}, 
  title={{QULATIS}: A Quantum Error Correction Methodology toward Lattice Surgery}, 
  year={2022},
  volume={},
  number={},
  pages={274-287},
  doi={10.1109/HPCA53966.2022.00028}
}

@article{Dennis2002,
	title = {Topological quantum memory},
	author = {Dennis, Eric and Kitaev, Alexei and Landahl, Andrew and Preskill, John},
	year = 2002,
	month = sep,
	publisher = {American Institute of Physics},
	journal = {Journal of Mathematical Physics},
	volume = {43},
	number = {9},
	pages = {4452--4505},
	doi = {10.1063/1.1499754},
}

@inproceedings{knapen2026pinball,
   title={Pinball: A Cryogenic Predecoder for Quantum Error Correction Decoding Under Circuit-Level Noise},
   url={http://dx.doi.org/10.1109/HPCA68181.2026.11408464},
   DOI={10.1109/hpca68181.2026.11408464},
   booktitle={2026 IEEE International Symposium on High Performance Computer Architecture (HPCA)},
   publisher={IEEE},
   author={Knapen, Alexander and Tao, Guanchen and Mack, Jacob and Bruno, Tomas and Saligane, Mehdi and Sylvester, Dennis and Zhang, Qirui and Ravi, Gokul Subramanian},
   year={2026},
   month=Jan, pages={1--17} }

@ARTICLE{senior2026aq2,
       author = {{Senior}, Andrew W. and {Edlich}, Thomas and {Heras}, Francisco J.~H. and {Zhang}, Lei M. and {Higgott}, Oscar and {Spencer}, James S. and {Applebaum}, Taylor and {Blackwell}, Sam and {Ledford}, Justin and {{\v{Z}}emgulyt{\.{e}}}, Akvil{\.{e}} and {{\v{Z}}{\'\i}dek}, Augustin and {Shutty}, Noah and {Cowie}, Andrew and {Li}, Yin and {Holland}, George and {Brooks}, Peter and {Beattie}, Charlie and {Newman}, Michael and {Davies}, Alex and {Jones}, Cody and {Boixo}, Sergio and {Neven}, Hartmut and {Kohli}, Pushmeet and {Bausch}, Johannes},
        title = "{A scalable and real-time neural decoder for topological quantum codes}",
      journal = {arXiv e-prints},
         year = 2025,
        month = dec,
          eid = {arXiv:2512.07737},
        pages = {arXiv:2512.07737},
          doi = {10.48550/arXiv.2512.07737},
archivePrefix = {arXiv},
       eprint = {2512.07737},
 primaryClass = {quant-ph},
       adsurl = {https://ui.adsabs.harvard.edu/abs/2025arXiv251207737S}
}

@article{valentino2025quekuf,
    author = {Valentino, Federico and Branchini, Beatrice and Conficconi, Davide and Sciuto, Donatella and Santambrogio, Marco D.},
    title = {{QUEKUF}: An {FPGA} Union Find Decoder for Quantum Error Correction on the Toric Code},
    year = {2025},
    issue_date = {September 2025},
    publisher = {Association for Computing Machinery},
    address = {New York, NY, USA},
    volume = {18},
    number = {3},
    issn = {1936-7406},
    url = {https://doi.org/10.1145/3733239},
    doi = {10.1145/3733239},
    journal = {ACM Trans. Reconfigurable Technol. Syst.},
    month = aug,
    articleno = {33},
    numpages = {26}
}

@article{liyanage2024helios,
   title={{FPGA}-Based Distributed Union-Find Decoder for Surface Codes},
   volume={5},
   ISSN={2689-1808},
   url={http://dx.doi.org/10.1109/TQE.2024.3467271},
   DOI={10.1109/tqe.2024.3467271},
   journal={IEEE Transactions on Quantum Engineering},
   publisher={Institute of Electrical and Electronics Engineers (IEEE)},
   author={Liyanage, Namitha and Wu, Yue and Tagare, Siona and Zhong, Lin},
   year={2024},
   pages={1--18}
}

@INPROCEEDINGS{liyanage2025network,
  author={Liyanage, Namitha and Wu, Yue and Houghton, Emmet and Zhong, Lin},
  booktitle={2025 IEEE International Conference on Quantum Computing and Engineering (QCE)}, 
  title={Network-Integrated Decoding System for Real-Time Quantum Error Correction with Lattice Surgery}, 
  year={2025},
  volume={01},
  number={},
  pages={1148-1159},
  doi={10.1109/QCE65121.2025.00129}}

@misc{yang2026realtimesurfacecodeerrorcorrection,
      title={Real-time Surface-Code Error Correction Using an {FPGA}-based Neural-Network Decoder}, 
      author={Xiaohan Yang and Xuandong Sun and Zhiyi Wu and Jiawei Zhang and Ji Jiang and Xiayu Linpeng and Yuxuan Zhou and Ji Chu and Jingjing Niu and Youpeng Zhong and Song Liu and Dapeng Yu},
      year={2026},
      eprint={2605.04892},
      archivePrefix={arXiv},
      primaryClass={quant-ph},
      url={https://arxiv.org/abs/2605.04892}, 
}

@misc{yan2026rethinkroleneuraldecoders,
      title={Rethink the Role of Neural Decoders in Quantum Error Correction}, 
      author={Ge Yan and Shanchuan Li and Yuxuan Du},
      year={2026},
      eprint={2605.12046},
      archivePrefix={arXiv},
      primaryClass={quant-ph},
      url={https://arxiv.org/abs/2605.12046}, 
}

@misc{knapen2026mitigatingclassicalresourcecosts,
      title={Mitigating Classical Resource Costs in Quantum Error Correction via Generalized {qLDPC} Predecoding}, 
      author={Alexander Knapen and Junyi Luo and Guanchen Tao and Yuxuan Wang and Tomas Bruno and Qirui Zhang and Dennis Sylvester and Mehdi Saligane and Gokul Subramanian Ravi},
      year={2026},
      eprint={2605.03180},
      archivePrefix={arXiv},
      primaryClass={quant-ph},
      url={https://arxiv.org/abs/2605.03180}, 
}

@article{gidney_2021,
    title={How to factor 2048 bit RSA integers in 8 hours using 20 million noisy qubits},
    volume={5},
    ISSN={2521-327X},
    url={http://dx.doi.org/10.22331/q-2021-04-15-433},
    DOI={10.22331/q-2021-04-15-433},
    journal={Quantum},
    publisher={Verein zur Forderung des Open Access Publizierens in den Quantenwissenschaften},
    author={Gidney, Craig and Ekerå, Martin},
    year={2021},
    month=Apr, pages={433}
}

@misc{báscones2026scalablefpgaarchitecturerealtime,
      title={A Scalable {FPGA} Architecture for Real-Time Decoding of Quantum {LDPC} Codes Using {GARI}}, 
      author={Daniel Báscones and Arshpreet Singh Maan and Valentin Savin and Francisco Garcia-Herrero},
      year={2026},
      eprint={2605.01035},
      archivePrefix={arXiv},
      primaryClass={quant-ph},
      url={https://arxiv.org/abs/2605.01035}, 
}

\appendix

\section{Other Decoder Confidence Scores}
\label{sec:other_decoder_confidence_scores}
In this section,
we mention two other DCSs
with minimal impact on resource utilisation.
These are:
\begin{enumerate}
    \item \textbf{Drop Frequency:} How fast the decoder generates a correction; specifically, the reciprocal of the number of clock cycles between two consecutive drop states.
    The idea behind this is that
	an easy decoding problem
	generates no large clusters,
	and only large clusters require many clock cycles
	between consecutive drops.
    \item \textbf{Minimum Defect Height:} The height of the lowest (unannihilated) defect in the decoding window.
    The idea behind this is that
	for easy decoding problems,
	the clusters are inactivated (i.e.\ resolved)
	at the top of the decoding window.
	A defect that is low in the window
	implies an active (i.e.\ unresolved) cluster that has grown a lot,
	indicating a difficult decoding problem.
\end{enumerate}
We evaluated the accuracy of all three DCSs
via Monte Carlo simulation,
and observe the cluster size 1-norm fraction to
significantly outperform the other two.

\section{Additional Active Depth Plot}
\label{sec:additional_active_depth_plot}
\begin{figure}[h]
    \centering
    \includegraphics[width=0.8\columnwidth]{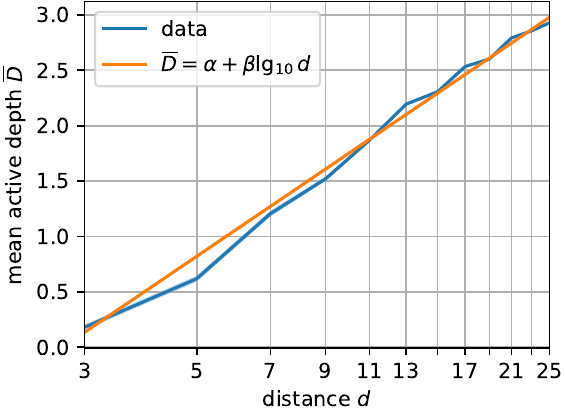}
    \caption{The same data as in \zcref{fig:active_depth},
    but focusing on the mean of each distribution;
    the standard error is thinner than the line width.
    The logarithmic fit (orange) parameters are
    $\alpha =-1.3(1)$ and $\beta =3.09(9)$.}
    \label{fig:mean_active_depth}
\end{figure}

\section{Pseudocode for the 2D Architecture}
\label{sec:pseudocode_for_the_2d_architecture}
In this section
we provide pseudocode for the 2D architecture of Snowflake
proposed in \zcref{sec:arch_2d}.
For readers unfamiliar with the
high-level idea of how Snowflake works
(i.e.\ how cluster grow, touch, merge,
and how defects are paired within them),
we recommend reading \cite{chan2026snowflake}
up to and including \S4.

This pseudocode is for the 2:1 cluster growth schedule
rather than the 1:1 schedule~\cite[\S 4.2]{chan2026snowflake}.
Much of this is similar to \cite[Appendix~D]{chan2026snowflake}
but with some differences.
In \zcref{sec:setup} we list the variables each component stores
and \zcref[range]{sec:high_level_procedures,sec:processes_common_to_both_merging_stages}
we explain how they are manipulated
in order to effect Snowflake.

\subsection{Setup}\label{sec:setup}

Denote the decoding window as $W$,
and its node and edge sets as $V_W$ and $E_W$, respectively.
We begin by associating each node $v \in V_W$
with a coordinate $(x, y, D)$
where $(x, y)$ is the spatial coordinate of the stabiliser in the surface code,
and $D$ is the depth of the node within $W$.
A \emph{sheet} is a set of nodes with the same depth,
and a \emph{column} is a set of nodes with the same spatial coordinate.

\begin{figure}[H]
	\centering
	\begin{tikzpicture}[
		stage/.style={
			draw=black,
			text centered,
			text depth=.25ex,
			text height=1.5ex
		},
		arrow/.style={-{Stealth[length=2mm]}},
	]
	\node[stage] (drop) {\texttt{drop}};
	\node[stage] (grow_whole) [right=of drop] {\texttt{grow\_whole}};
	\node[stage] (merging_whole) [right=of grow_whole] {\texttt{merging\_whole}};
	\node[stage] (grow_half) [below=of grow_whole] {\texttt{grow\_half}};
	\node[stage] (merging_half) [below=of merging_whole] {\texttt{merging\_half}};
	\draw[arrow] (drop.east) -- (grow_whole.west);
	\draw[arrow] (grow_whole.east) -- (merging_whole.west);
	\draw[arrow] ([xshift=2mm]merging_whole.south) -- ++(0,-4mm) -- ++(15mm,0) |- (merging_whole.east);
	\draw[arrow] (merging_whole.south west) -- (grow_half.north east);
	\draw[arrow] (grow_half.east) -- (merging_half.west);
	\draw[arrow] ([xshift=2mm]merging_half.south) -- ++(0,-4mm) -- ++(15mm,0) |- (merging_half.east);
	\draw[arrow] ([xshift=-2mm]merging_half.south) -- ++(0,-4mm) -| (drop.south);
	\end{tikzpicture}
	\caption{Flowchart for the stages of Snowflake:
	\texttt{drop}, \texttt{grow\_whole}, and \texttt{grow\_half} are of fixed duration
	whereas the other two last a variable duration,
	hence their loops.}
	\label{fig:stage_flowchart}
\end{figure}
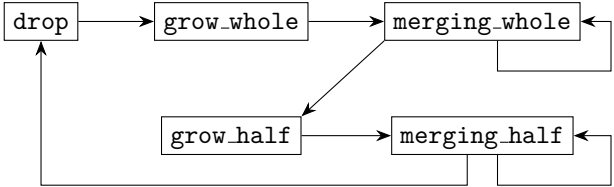

As \zcref{fig:decoder_sliced} shows,
the architecture comprises a 2D grid of PEs,
a global controller,
and memory, which we discuss in more detail:
\begin{itemize}
    \item Each PE corresponds to one column of nodes
    (see \zcref{fig:2d_architecture}),
    and can modify the variables of three consecutive nodes
    within the column at any given time.
    \item The controller can broadcast the same boolean to,
    and receive booleans from,
    every PE.
    \item The memory stores all the variables
    introduced in \zcref{sec:node_variables,sec:edge_variables}.
\end{itemize}
At a high level,
Snowflake can be thought of as operating
across discretised timesteps
(or \emph{generations} in cellular-automata language).
In any given timestep,
Snowflake is in one of the five stages,
following the flowchart in \zcref{fig:stage_flowchart}.
In other words,
we define the following.
\begin{defn}\label{defn:timestep}
A \emph{timestep} is a
transition along one of the arrows in \zcref{fig:stage_flowchart}.
\end{defn}
\noindent
As for hardware,
a timestep would equate to multiple clock cycles.
\begin{defn}\label{defn:decoding_cycle}
A \emph{decoding cycle} is defined as
one cycle through all five stages
i.e.\ the series of transitions between successive \texttt{drop} stages.
\end{defn}
\noindent
The decoding cycle relates to hardware in that
its duration is precisely the time taken
to process one stabiliser measurement round.

At a high level, the following processes occur in each stage.
In \texttt{drop},
the decoding window $W$ slides up the decoding graph by one layer.
In \texttt{grow\_whole},
a specific subset of active clusters in $W$ grow outward by half an edge,
and in \texttt{grow\_half},
the complementary set of active clusters grow.
In both merging stages,
newly touching clusters merge
and Snowflake updates
(1) whether each cluster is active or inactive,
and (2) the defect pairings within each cluster.

\subsubsection{Node Variables}\label{sec:node_variables}
Each node $v \in V_W$ is assigned a unique integer \verb|ID|
and possesses the following variables:

\begin{enumerate}
	\item \verb|active| is a boolean indicating
	whether $v$ is in an active cluster.
	If $v.\verb|active| =\verb|true|$ we say $v$ is active.
	\item \texttt{whole} is a boolean indicating
	whether $v$ is in a whole cluster.
	If $v.\texttt{whole} =\texttt{true}$ we say $v$ is whole.
	\item \verb|CID| is an integer indicating
	the cluster that $v$ belongs to.
	Clusters are identified by the lowest \verb|ID| of all its nodes.
	The node of this \verb|ID| is the root.
	\verb|CID| can also adopt the value \verb|reset|
	to indicate $v$ should \emph{unroot}
	(defined in \zcref{sec:unrooting})
	in the next timestep (\zcref{defn:timestep}).
	\item \verb|defect| is a boolean indicating whether $v$ has a defect.
	This can be thought of as a particle which is passed between nodes.
	When a node receives two defects in a timestep,
	they annihilate.
	\item \verb|pointer| is the direction $v$ should push the defect
	(by flipping the corresponding edge).
	Possible values (under circuit-level noise) are
		\texttt{C},
		\texttt{N},
		\texttt{W},
		\texttt{E},
		\texttt{S},
		\texttt{D},
		\texttt{U},
		\texttt{NU},
		\texttt{WD},
		\texttt{EU},
		\texttt{SD},
		\texttt{NWD},
		\texttt{SEU},
	representing respectively
		centre,
		north,
		west,
		east,
		south,
		down,
		up,
	and combinations thereof.
	Following the pointers starting from $v$ should eventually lead to
	the root of the cluster that $v$ is in.
	Only roots have $\verb|pointer| =\verb|C|$
	as they do not push defects but accumulate them.
	\item \texttt{grown} is a boolean indicating
	whether the cluster that $v$ is in
	has grown in the first growth round of the current decoding cycle
    (\zcref{defn:decoding_cycle}).
	\item \verb|unrooted| is a boolean indicating whether
	$v$ has unrooted in the current decoding cycle.
	\item \verb|busy| is a boolean indicating whether $v$ is busy.
\end{enumerate}

\begin{figure*}
	\centering
	\includegraphics[width=0.7\textwidth]{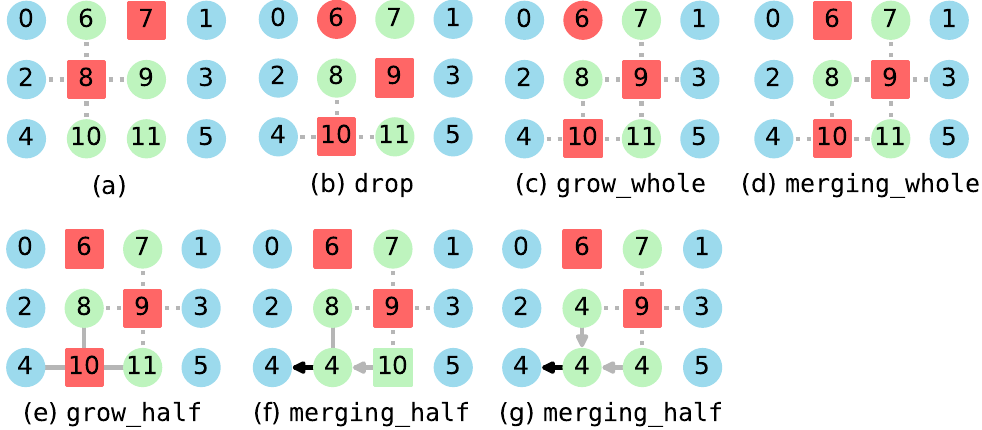}
	\caption{An example of a \emph{decoding cycle} (\zcref{defn:decoding_cycle})
    that takes 6 \emph{timesteps} (\zcref{defn:timestep}),
    for a distance-3 repetition code.
    Ungrown edges are invisible;
	half-grown, dotted;
	fully grown, solid.
	Edges flipped by the decoder are in black.
	Active nodes are squares;
	inactive nodes, circles.
	\texttt{CID} is shown as a label.
	Pointers are shown by arrows on edges.
	(a) Each node is in its own cluster
	so has $\texttt{CID} =\texttt{ID}$.
	Two clusters are active.
	(b) \emph{All} relative variables fall by one layer,
	including \texttt{CID} which is adjusted to refer to the node below.
	A new defect appears at the top.
	(c) Node 9 grows its incident edges by $\frac12$
	as it is active and whole.
	(d) Node 6 recognises it is a root with a defect,
	so becomes active.
	(e) Node 10 grows its incident edges by $\frac12$
	as it is active and half.
	Nodes 10 and 11 now each see a lower \texttt{CID} than their own
	when they look at their neighbours along fully grown edges.
	(f) They adopt that \texttt{CID}
	and point toward that neighbour;
	thus the defect at node 10 is pushed along its pointer
	to the root, which flips an edge.
	Meanwhile, node 8 now sees a lower \texttt{CID} than its own.
	(g) Node 8 updates its \texttt{CID} and \texttt{pointer},
	so now the 4-node cluster is fully flooded.
	Throughout this stage,
	each non-root node was adopting the \texttt{active} variable
	of the node it was pointing to.
	This activity propagation also completes in this timestep,
	so the 4-node cluster is fully synced.}
	\label{fig:drop_grow_merging}
\end{figure*}

\noindent
We refer to the first five variables as \emph{relative}
and the last three as \emph{absolute}.

\subsubsection{Edge Variables}\label{sec:edge_variables}
Each edge $e \in E_W$ has the following relative variables:
\begin{enumerate}
	\item \verb|growth| is its growth value
	(how much of that edge is contained in any cluster)
	which is in $\{0, \frac12, 1\}$.
	\item \verb|correction| is a boolean indicating
	if the decoder (tentatively) thinks $e$ should be flipped
    (but this can change later on).
\end{enumerate}
These edge variables are stored in a lookup table
(a global register array implemented using FPGA flip-flops,
as in the main text).

\subsection{High-Level Procedures}\label{sec:high_level_procedures}

The controller repeatedly runs \zcref{alg:controller_cycle};
this follows the flowchart in \zcref{fig:stage_flowchart}.
\begin{algorithm}[H]
\caption{Run by controller repeatedly.}\label{alg:controller_cycle}
\begin{algorithmic}
	\State \Call{Drop}{}
    \For{$\texttt{metastage} =(\texttt{whole}, \texttt{half})$}
        \State \Call{Conduct}{\textsc{Grow}[\texttt{metastage}], $D$}
        \Repeat
            \State $D \gets$ active depth
			\Comment{See \zcref{defn:active_depth}.}
            \State \Call{Conduct}{\textsc{Merging}[\texttt{metastage}], $D +1$}
        \Until{not any $v.\texttt{busy}~\forall v \in V_W$}
    \EndFor
\end{algorithmic}
\end{algorithm}
\noindent
The \textsc{Drop} and \textsc{Conduct} procedures are defined
in \zcref{alg:controller_procedures},
with one of them being called per timestep (\zcref{defn:timestep}).
Stage \texttt{drop} is handled entirely by the controller,
and its effect will be described in \zcref{sec:fixed_duration_stages}.
\textsc{Conduct} is simply a wrapper
that takes in a node procedure from \zcref{alg:node_procedures}
and an integer $D \in \{0, \dots, h\}$ representing the current active depth,
then performs one timestep of Snowflake.
The early-stop feature is implemented here:
during the growth (merging) stages,
processing is skipped for nodes deeper (one deeper than) $D$.

In the next two subsections we explain the effect of each stage.
Though the pseudocode
explains exactly what each component does,
it is difficult to infer from this the emergent effect,
so we describe the latter in the accompanying prose.

\begin{algorithm}[H]
\caption{The two procedures that can be called in \zcref{alg:controller_cycle}.
\textsc{Proc} is a procedure from \zcref{alg:node_procedures}.}\label{alg:controller_procedures}
\begin{algorithmic}[2]
    \zcsetup{reftype=line}
	\Procedure{Drop}{}
		\For{$e \in \text{bottom layer of $W$}$}
			\If{$e.\texttt{correction}$}
			\Comment{Commit $e$.}
				\State add $e$ to the correction
			\EndIf
		\EndFor
        \For{$v \in V_W$}
            \State $v.\texttt{grown} \gets \texttt{false}$ \label{line:reset_grown}
            \State $v.\texttt{CID} \gets$ \texttt{ID} of node below current root
        \EndFor
        \State shift all relative variables down one layer
		\For{$v \in$ top sheet of $W$}
		\LComment{Inherit the new syndrome sheet.}
			\State $v.\texttt{defect} \gets$ whether
			the measurement of the corresponding ancilla qubit
			differs from the previous round \label{line:set_defect}
			\State $v.\texttt{active} \gets \texttt{false}$ \label{line:reset_active}
		\EndFor
	\EndProcedure
	\State
    \Procedure{Conduct}{\textsc{Proc}, $D \in \{0, \dots, h\}$}
        \For{$L =1, \dots, D$}
            \State load node sheets $L-1, L, L+1$ into 2D grid
            \ForAll{$v$ at depth $L$}
                \State \Call{Proc}{$v$}
            \EndFor
        \EndFor
	\EndProcedure
\end{algorithmic}
\end{algorithm}

\subsection{Fixed-Duration Stages}\label{sec:fixed_duration_stages}
As mentioned in the previous subsection,
the \texttt{drop} stage is handled entirely by the controller.
The controller adds to the correction
(i.e.\ finalises the flipped-or-not-flipped status
of all edges in the lowest layer of $W$)
and raises $W$ by one layer,
as in \zcref{fig:drop_grow_merging}b.
The latter can be done simply by updating the memory addresses of the nodes rather than physically moving the data.
Each node $v$ in the top sheet of $W$ starts in its own cluster
\cite[Definition~8]{chan2026snowflake},
thus its relative variables are set to reflect this
(\zcref{alg:controller_procedures} \zcref{line:set_defect,line:reset_active}).
$v.\texttt{whole}$,
$v.\texttt{CID}$,
and $v.\texttt{pointer}$
need not be explicitly reset
as they never change from their correct values of
\texttt{true},
$v.\texttt{ID}$,
and \texttt{C},
respectively.

\begin{algorithm}[H]
\caption{The four procedures that can be called in \zcref{alg:controller_procedures}.
\texttt{metastage} can be \texttt{whole} or \texttt{half}.}\label{alg:node_procedures}
\begin{algorithmic}[2]
    \zcsetup{reftype=line}
	\Procedure{Grow}{}[$\texttt{whole}$]($v$)
		\If{$v.\texttt{active}$ and $v.\texttt{whole}$}
			\State \Call{SubGrow}{$v$}
			\State $v.\texttt{grown} \gets \texttt{true}$
		\EndIf
	\EndProcedure
	\State
	\Procedure{Grow}{}[$\texttt{half}$]($v$)
		\State $v.\texttt{unrooted} \gets \texttt{false}$ \label{line:reset_unrooted}
		\If{$v.\texttt{active}$ and not $v.\texttt{whole}$ and not $v.\texttt{grown}$}
			\State \Call{SubGrow}{$v$}
		\EndIf
	\EndProcedure
	\State
	\Procedure{Merging}{}[$\texttt{metastage}$]($v$)
		\State $v.\texttt{busy} \gets \texttt{false}$
		\State \Call{Flooding}{$v, \texttt{metastage}$}
		\State \Call{Syncing}{$v$}
	\EndProcedure
\end{algorithmic}
\end{algorithm}

Now referring to \zcref{alg:node_procedures}
and the subroutine \textsc{SubGrow} in \zcref{alg:subroutines},
stage \texttt{grow\_whole} grows all edges around
each active whole cluster by $\frac12$,
as in \zcref{fig:drop_grow_merging}c.
Stage \texttt{grow\_half} does the same
but for active half clusters that have not already grown
in the current decoding cycle (\zcref{defn:decoding_cycle}),
as in \zcref{fig:drop_grow_merging}e.

\subsection{Variable-Duration Stages}
\label{sec:processes_common_to_both_merging_stages}
\zcref[S]{fig:drop_grow_merging}d shows a merging example
that lasts just one timestep (\zcref{defn:timestep}),
and f--h shows one that needs three timesteps.
The \texttt{merging\_whole} stage is the same as
\texttt{merging\_half} but with one extra process,
so we first explain the latter:
it comprises two processes,
\emph{flooding} and \emph{syncing},
which occur independently.
We thus split \textsc{Merging} into two subroutines,
given in \zcref{alg:subroutines},
effecting each process.

The processes in \zcref[range]{sec:flooding,sec:unrooting}
are common to both merging stages
and references to specific lines refer to those in \zcref{alg:subroutines}.
The process in \zcref{sec:processes_unique_to_merging_whole}
is unique to the \texttt{merging\_whole} stage.

\subsubsection{Flooding}\label{sec:flooding}
In flooding,
newly touching clusters adopt the lowest \verb|CID| among them.
This \verb|CID| propagates as a flood
starting from where they touched:
each node looks at its neighbours along fully grown edges.
If it sees a lower \verb|CID| than its own,
it adopts that \verb|CID|
and points toward that neighbour
(\zcref[range]{line:see_lower_CID,line:match_CID}).
After this process,
all nodes in a given cluster will have the same \verb|CID|,
and following the pointers
will always lead to the root promised by \verb|CID|.

\begin{rem}\label{rem:boundaries_lower_ID_than_detectors}
Crucial to the algorithm is that
boundary nodes are of lower \verb|ID| than detectors.
Each root can thus determine if its cluster touches a boundary
by checking if itself is a boundary node.
\end{rem}

\subsubsection{Syncing}
In syncing,
all defects are pushed along pointers toward roots,
one edge per timestep (\zcref[range]{line:if_x,line:flip_uv_correction}).
Eventually,
each root will know whether its cluster is active (\zcref{line:active_gets_x}),
using the following lemma from \cite[\S A.2]{Chan2023c}.
\begin{lem}
A cluster is inactive iff it has an even defect count
or touches a boundary.
\end{lem}
\noindent
At the same time,
this activity knowledge propagates
in the opposite direction
from root to the rest of the cluster (\zcref{line:get_active_from_pointee}).
After this process,
a node is in an active cluster iff it is active.

\subsubsection{Unrooting}\label{sec:unrooting}
\begin{figure}
	\centering
	\includegraphics[width=0.48\textwidth]{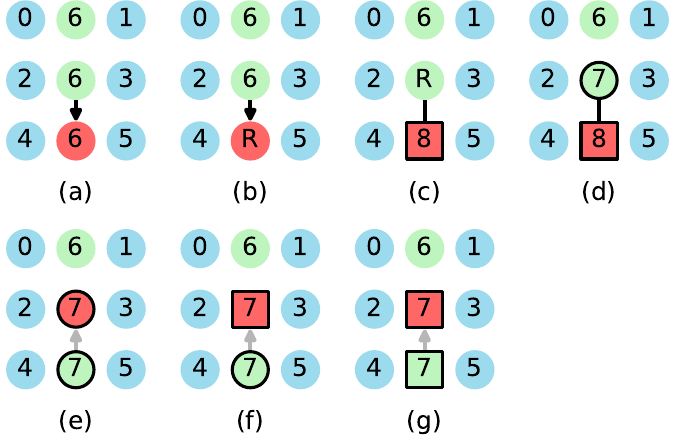}
	\caption{A demonstration of unrooting on
	the distance-2 repetition code.
	Relative variables are shown as in \zcref{fig:drop_grow_merging}.
	\texttt{CID} value \texttt{reset} is labelled `\textsf{R}'.
	Nodes with \texttt{unrooted} as \texttt{true} are outlined in black.
	(a) There is a 2-node cluster
	comprising Nodes 7 (the green one) and 8 (the red one).
	Node 8 has a downward pointer (not shown).
	(b) Thus, Node 8 starts unrooting.
	(c) Node 7 sees this signal to unroot
	so starts unrooting.
	Node 8 finishes unrooting.
	(d) Node 7 finishes unrooting.
	(e--g) \texttt{merging\_whole} completes.}
	\label{fig:unrooting}
\end{figure}
Unrooting is a process needed when
a downward pointer is dropped onto
a node in the bottom sheet of $W$
(can happen while a cluster falls out of view),
and that node receives a defect.
This breaks the pointer tree structure established in the previous decoding cycle
(\zcref{defn:decoding_cycle}),
as $\exists$ nonempty $S \subset V_W$
from which following the pointers will lead past the bottom of $W$.
To fix it,
each node in $S$ resets its \verb|CID| and \verb|pointer|
so that the structure can be rebuilt from scratch,
via further \texttt{merging\_whole} or \texttt{merging\_half}
timesteps.
\zcref[S]{fig:unrooting} shows an example.

The breaking point is found in \zcref{line:pointer_D}
whence the signal to unroot,
$\verb|CID| =\verb|reset|$,
propagates during \texttt{merging\_whole} or \texttt{merging\_half}
as a flood through the cluster;
each node unroots itself as soon as it receives this signal
(\zcref[range]{line:access,line:break}).
It takes 2 timesteps for a node to unroot,
after which it cannot unroot again in the same decoding cycle
(\zcref{line:if_not_unrooted}).
This ensures the flood does not backpropagate.
The memory of whether a node has unrooted
resets after this stage finishes (\zcref{alg:node_procedures} \zcref{line:reset_unrooted}).

The higher up a node is in $W$,
the lower its \verb|ID|;
this is for two reasons.
First,
it makes unrooting less likely
as pointer paths tend to lead upward rather than downward.
Second,
it ensures defects in active clusters
are pushed to a highest point
so they can annihilate with defects in newer, smaller clusters
as early as possible.

\subsubsection{The Process Unique to \texttt{merging\_whole}}
\label{sec:processes_unique_to_merging_whole}
The process unique to the \texttt{merging\_whole} stage
is tracking which clusters have grown.
The variable \texttt{grown} floods
in the same fashion as \texttt{cid}
but with no connection to \texttt{pointer}
(\zcref{alg:subroutines} \zcref[range]{line:see_grown_true,line:set_grown_to_true}).
This ensures all nodes in a given cluster
will have the same boolean for \texttt{grown}.
The memory of whether a cluster has grown
resets after each decoding cycle (\zcref{alg:node_procedures} \zcref{line:reset_grown}).

\begin{figure*}
\begin{minipage}{\linewidth}
\begin{algorithm}[H]
\caption{Subroutines for \zcref{alg:node_procedures}.}\label{alg:subroutines}
\begin{algorithmic}[2]
    \zcsetup{reftype=line}
	\Procedure{SubGrow}{$v$}
		\For{each edge $e$ incident to $v$}
			\State $e.\texttt{growth} \gets \min\{1, e.\texttt{growth} +\frac12\}$
		\EndFor
		\State flip $v.\texttt{whole}$
	\EndProcedure
	\State
	\Procedure{Flooding}{$v, \texttt{metastage} \in \{\texttt{whole}, \texttt{half}\}$}
		\If{$v.\texttt{CID} =\texttt{reset}$}
		\Comment{Finish unrooting $v$.}
			\State $v.\texttt{busy} \gets \texttt{true}$
			\State $v.\texttt{CID} \gets v.\texttt{ID}$
			\State $v.\texttt{unrooted} \gets \texttt{true}$
		\Else
			\ForAll{$u: uv \in E_W \wedge uv.\texttt{growth} =1$} \label{line:access}
				\If{$u.\texttt{CID} =\texttt{reset}$}
					\If{not $v.\texttt{unrooted}$} \label{line:if_not_unrooted}
						\State \Call{StartUnroot}{$v$}
						\State break \label{line:break}
					\EndIf
				\ElsIf{$u.\texttt{CID} <v.\texttt{CID}$} \label{line:see_lower_CID}
                    \State $v.\texttt{busy} \gets \texttt{true}$
                    \State $v.\texttt{pointer} \gets$ toward $u$
                    \State $v.\texttt{CID} \gets u.\texttt{CID}$ \label{line:match_CID}
				\EndIf
				\If{$\texttt{metastage} =\texttt{whole}$ and $u.\texttt{grown}$ and not $v.\texttt{grown}$} \label{line:see_grown_true}
					\State $v.\texttt{busy} \gets \texttt{true}$
					\State $v.\texttt{grown} \gets \texttt{true}$ \label{line:set_grown_to_true}
				\EndIf
			\EndFor
		\EndIf
	\EndProcedure
    \State
	\Procedure{Syncing}{$v$}
		\State $x :=(v~\text{is a detector})$ and $v.\texttt{defect}$
		\If{$v.\texttt{pointer} =\texttt{C}$}
		\Comment{$v$ is a root.}
			\State $v.\texttt{active} \gets x$ \label{line:active_gets_x}
		\ElsIf{$v \in$ bottom sheet of $W$ and
		$v.\texttt{pointer} \in\{
			\texttt{D},
			\texttt{WD},
			\texttt{SD},
			\texttt{NWD}
		\}$}
            \If{$x$} \label{line:pointer_D}
                \State \Call{StartUnroot}{$v$}
            \EndIf
		\Else
			\State $u :=$ node toward $v.\texttt{pointer}$
			\State $v.\texttt{active} \gets u.\texttt{active}$ \label{line:get_active_from_pointee}
			\If{$x$} \label{line:if_x}
				\State $v.\texttt{busy} \gets \texttt{true}$
				\State flip $uv.\texttt{correction}$ to push defect to $u$\label{line:flip_uv_correction}
			\EndIf
		\EndIf
		\If{$v.\texttt{active}$ changed}
			\State $v.\texttt{busy} \gets \texttt{true}$
		\EndIf
	\EndProcedure
	\State
    \Procedure{StartUnroot}{$v$}
        \State $v.\texttt{busy} \gets \texttt{true}$
        \State $v.\texttt{CID} \gets \texttt{reset}$
        \State $v.\texttt{pointer} \gets \texttt{C}$
	\EndProcedure
\end{algorithmic}
\end{algorithm}
\end{minipage}
\end{figure*}

\end{document}